\documentclass[aps,twocolumn,superscriptaddress,preprintnumbers]{revtex4}

\usepackage[latin9]{inputenc}
\usepackage{color}
\usepackage{verbatim}
\usepackage{amsmath}
\usepackage{amssymb}
\usepackage{graphicx}
\usepackage{esint}
\usepackage{hyperref}

\makeatletter
\@ifundefined{textcolor}{}
{%
 \definecolor{BLACK}{gray}{0}
 \definecolor{WHITE}{gray}{1}
 \definecolor{RED}{rgb}{1,0,0}
 \definecolor{GREEN}{rgb}{0,1,0}
 \definecolor{BLUE}{rgb}{0,0,1}
 \definecolor{CYAN}{cmyk}{1,0,0,0}
 \definecolor{MAGENTA}{cmyk}{0,1,0,0}
 \definecolor{YELLOW}{cmyk}{0,0,1,0}
}

\usepackage{mathrsfs}

\draft
\makeatother

\begin{document}

\title{Non-Abelian Majorana Doublets in Time-Reversal Invariant Topological Superconductors}

\author{Xiong-Jun Liu \footnote{email: phyliuxiongjun@gmail.com}}
\affiliation{Department of Physics, Hong Kong University of Science and Technology, Clear Water Bay, Hong Kong, China}
\affiliation{Institute for Advanced Study, Hong Kong University of Science and Technology, Clear Water Bay, Hong Kong, China}
\author{Chris L. M. Wong}
\affiliation{Department of Physics, Hong Kong University of Science and Technology, Clear Water Bay, Hong Kong, China}
\author{K. T. Law \footnote{email: phlaw@ust.hk}}
\affiliation{Department of Physics, Hong Kong University of Science and Technology, Clear Water Bay, Hong Kong, China}

\begin{abstract}
The study of non-Abelian Majorana zero modes advances our understanding of the fundamental physics in quantum matter, and pushes the potential applications of such exotic states to topological quantum computation. It has been shown that in two-dimensional (2D) and 1D chiral superconductors, the isolated Majorana fermions obey non-Abelian statistics. However, Majorana modes in a $Z_2$ time-reversal invariant (TRI) topological superconductor come in pairs due to Kramers' theorem. Therefore, braiding operations in TRI superconductors always exchange two pairs of Majoranas. In this work, we show interestingly that, due to the protection of time-reversal symmetry, non-Abelian statistics can be obtained in 1D TRI topological superconductors and may have advantages in applying to topological quantum computation. Furthermore, we unveil an intriguing phenomenon in the Josephson effect, that the periodicity of Josephson currents depends on the fermion parity of the superconducting state. This effect provides direct measurements of the topological qubit states in such 1D TRI superconductors.
\end{abstract}
\date{\today }
\maketitle

\indent

\section{Introduction}


The search for exotic non-Abelian quasiparticles has been a focus of both theoretical and experimental studies in condensed matter physics, driven by both the exploration of the fundamental physics and the promising applications of such modes to a building block for fault-tolerant topological quantum computer \cite{Moore,Read,Ivanov,Kitaev1,Kitaev2,Nayak,Sankar}. Following this pursuit, the topological superconductors have been brought to the forefront for they host exotic zero energy states known as Majorana fermions \cite{Fu1,Wilczek,Sau0,Alicea,Roman0,Oreg,Potter,Franz,Alicea1}. For two-dimensional (2D) chiral $p+ip$ pairing state, which breaks time-reversal symmetry, one Majorana mode exists in each vortex core \cite{Read}, and for 1D $p$-wave case, such state is located at each end of the system \cite{Kitaev1}. Due to the particle-hole symmetry, Majorana fermions in a topological superconductor are self-hermitian modes which are identical to their own antiparticles. A complex fermion, whose quantum states span the physical space in the condensed matter system, is formed by two Majoranas that can be located far away from each other. This allows to encode quantum information in the non-local fermionic states, which are topologically stable against local perturbations. Existence of $2n$ Majorana zero modes leads to $2^{n-1}$-fold ground-state degeneracy, and braiding two of such isolated modes in 2D or 1D superconductors transforms one state into another which defines the non-Abelian statistics \cite{Ivanov,Alicea1}. Remarkably, Majorana end states have been suggestively observed through tunneling measurements \cite{Law,Flensberg,Liu0} in 1D effective $p$-wave superconductors obtained by semiconductor nanowire/$s$-wave superconductor heterostructures \cite{Kouwenhoven,Deng,Das}.

Recently, a new class of topological superconductors with time-reversal symmetry, referred to as DIII symmetry class superconductor and classified by $Z_2$ topological invariant \cite{Ryu,Qi1,Teo, Schnyder, Beenakker}, have attracted rapidly growing efforts \cite{Qi1,Teo, Schnyder, Beenakker,Ortiz, Law2, Nagaosa, Kane, Sho}. Different from chiral superconductors, in DIII class superconductor the zero modes come in pairs due to Kramers' theorem. Many interesting proposals have been studied to realize $Z_2$ time-reversal invariant (TRI) Majorana quantum wires using proximity effects of $d$-wave, $p$-wave, $s\pm$-wave, or conventional $s$-wave superconductors. It was shown that at each end of such a quantum wire are localized two Majorana fermions which form a Kramers doublet and are protected by time-reversal symmetry \cite{Law2,Kane,Sho,Keselman}.

With the practicability in realization, a fundamental question is that can the DIII class topological superconductor be applied to topological quantum computation? The puzzle arises from the fact that braiding the end states in a DIII class 1D superconductor always exchanges Majorana Kramers pairs rather than isolated Majorana modes. While braiding two pairs of Majoranas in chiral topological superconductors yields Abelian operations, in this work, we show interestingly that braiding Majorana end states in DIII class topological superconductors is non-Abelian due to the protection of time-reversal symmetry. We further unveil an intriguing phenomenon in the Josephson effect, that the periodicity of Josephson currents depends on the fermion parity of the superconducting state, which provides direct measurements of all topological qubit states in the DIII class 1D superconductors.

The article is organized as follows. In Sec.~II, we briefly introduce how to engineer the DIII class 1D topological superconductor by inducing $p$-wave superconductivity in a conducting wire in proximity to a non-centrosymmetric superconductor. Then in Sec.~III, we turn to a detailed study of the non-Abelian Majorana doublets in the DIII class 1D topological superconductor. Section IV is devoted to investigate the Josephson effect, which shows an interesting strategy to read out topological qubit states in TRI superconductors. Finally the conclusions are given in Sec.~V.

\section{Topological superconductor of DIII class by proximity effect}

\begin{figure}[ht]
\includegraphics[width=0.8\columnwidth]{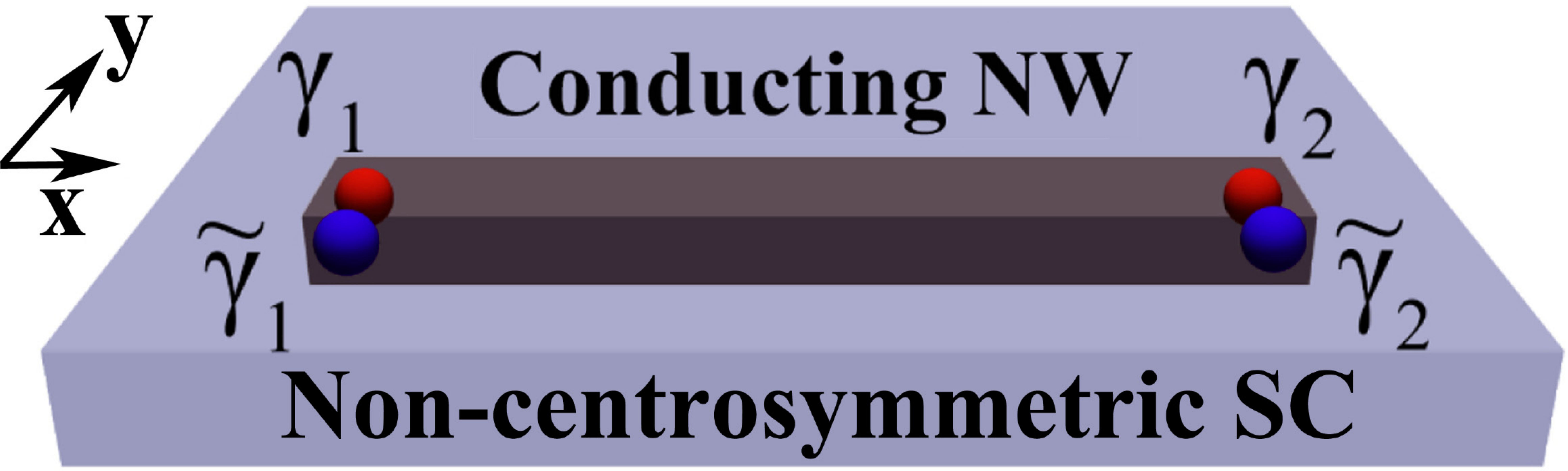}\caption{The DIII class 1D topological superconductor realized by depositing a conducting nanowire (NW) on the top of a non-centrosymmetric superconductor (SC). Both $s$- and $p$-wave pairings can be induced in the conducting wire by tunneling couplings through the interface between the NW and substrate superconductor.}
\label{proximity}
\end{figure}

Several interesting proposals have been considered to realize DIII class 1D topological superconductors, including to use proximity effects of $d$-wave, $p$-wave, and $s\pm$-wave superconductors~\cite{Law2,Kane,Sho,Keselman}. Here we briefly study how to engineer such $Z_2$ topological superconductor by
depositing a conducting quantum wire on a non-centrosymmetric superconductor thin-film which can induce $s$- and $p$-wave pairings in the wire by proximity effect \cite{Sigrist,Sato}, as illustrated in Fig.~\ref{proximity}.
The total Hamiltonian of the heterostructure system reads $H=H_{\rm SC}+H_{\rm wire}+H_{\rm t}$, where $H_{\rm SC}, H_{\rm wire}$, and $H_{\rm t}$ represent the Hamiltonians for the substrate superconductor, the conducting wire, and the tunneling at the interface, respectively. Due to the lack of inversion symmetry, a non-centrosymmetric superconductor has both $s$-wave and $p$-wave pairings \cite{Sigrist}. For convenience, we denote by the pairings in the substrate superconductor $\Delta^{(0)}_s$ and $\Delta^{(0)}_p$, respectively. The BdG Hamiltonian for the 2D non-centrosymmetric superconductor is given by
\begin{eqnarray}
 H_{\rm SC}&=&\sum_{k_x,k_y}\bigr[\epsilon_0(k_x,k_y)\tau_z +\alpha^{(0)}_R \sin k_y\sigma_x -\alpha^{(0)}_R \sin k_x\sigma_y\tau_z\nonumber\\
 &&+ \Delta_p^{(0)} \sin k_y\sigma_z  \tau_x -\Delta_p^{(0)} \sin k_x    \tau_y -\Delta_s^{(0)}  \sigma_y  \tau_y\bigr],
\end{eqnarray}
where $\epsilon_0(k_x,k_y)=-2t^{(0)}(\cos k_x + \cos k_y) -\mu^{(0)}$ is the normal dispersion relation with $t^{(0)}$ the hopping coefficient in the superconductor, $\sigma_j$ and $\tau_j$ ($j=x,y,z$) are the Pauli matrices acting on the spin and Nambu spaces, respectively, $\alpha^{(0)}_R$ is the spin-orbit coupling coefficient, and $\mu^{(0)}$ is chemical potential. The pairing order parameters can be reorganized by $ \hat{\Delta} = (\Delta_s^{(0)} + {\bf d}\cdot  \sigma )(i \sigma_y)$, with the ${\bf d}$-vectors defined as ${\bf d} = \Delta_p^{(0)}(- \sin k_y , \sin k_x, 0)$.

A single-channel 1D conducting quantum wire, being put along the $x$ axis, can be described by the following Hamiltonian
\begin{eqnarray}
H_{\rm wire}=\sum_{k_x}(-2t_w\cos k_x-\mu_w)\tau_z,
\end{eqnarray}
with $t_w$ the hopping coefficient and $\mu_w$ the chemical potential in the wire. It is noteworthy that the intrinsic spin-orbit interaction is not needed to reach the TRI topological superconducting phase, while the proximity effect can induce an effective spin-orbit interaction in the nanowire. Now we give the tunneling Hamiltonian $H_{\rm t}$ for the interface. For simplicity we consider that at the interface the coupling between the substrate superconductor and the nanowire is uniform, and thus the momentum $k_x$ is still a good quantum number. Then the tunneling Hamiltonian can be written down as
\begin{eqnarray}
H_{\rm t}= -t_\perp\sum_{k_x,\sigma}c^\dag_{\sigma}(k_x) d_{i_{y0},\sigma}(k_x)+h.c.,
\end{eqnarray}
where $\sigma=\uparrow,\downarrow$ are the spin indices, $t_\perp$ denotes the tunneling coefficient between the nanowire and substrate superconductor, $c^\dag_{\sigma},c_{\sigma}$ and $d^\dag_{\sigma},d_{\sigma}$ are the creation and annihilation operators of electrons for the quantum nanowire and the superconductor, respectively. The site number $i_{y0}$ characterizes where the heterostructure is located on the $y$ axis in the non-centrosymmetric superconductor.

\begin{figure}[ht]
\includegraphics[width=1\columnwidth]{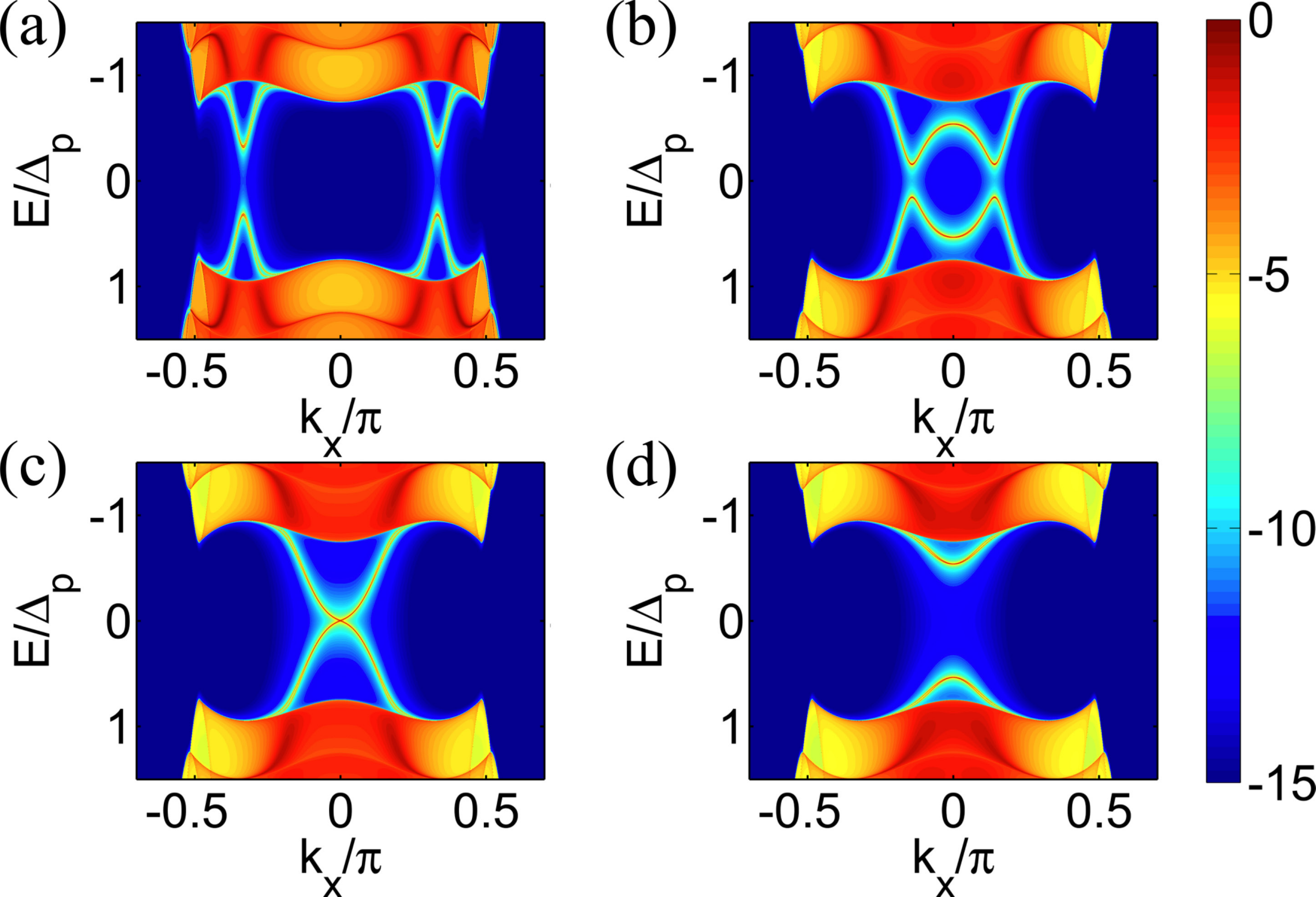}\caption{The logarithmic plot of the spectral function for the nanowire/non-centrosymmetric superconductor heterostructure. The yellow dotted curves show the bulk band structure of the nanowire system. The red solid areas (in the upper and lower positions of each panel) represent the bulk states of the substrate superconductor. (a) Topological regime with the chemical potential $\mu_w$ set as $-2t_w+5|\Delta_p^{(0)}|$ in the nanowire. In this regime at each end of the wire localized two Majorana zero modes [Fig.~\ref{SIboundwavefunction}]. (b) Topological regime with reduced bulk gap by tuning $\mu_w=-2t_w+|\Delta_p^{(0)}|$ close to the band bottom. (c) Critical point  $\mu_w^c=-2t_w$ for the topological phase transition with the bulk gap closed. (d) Trivial phase regime for the nanowire with  $\mu=-2t_w-|\Delta_p^{(0)}|$. Other parameters are taken that $t_\perp=0.5t_w=0.5t^{(0)}, |\Delta_s^{(0)}|=0.5|\Delta_p^{(0)}|$, and $\alpha_R^{(0)}=|\Delta_p^{(0)}|$.}
\label{SIspectrum}
\end{figure}
\begin{figure}[ht]
\includegraphics[width=0.9\columnwidth]{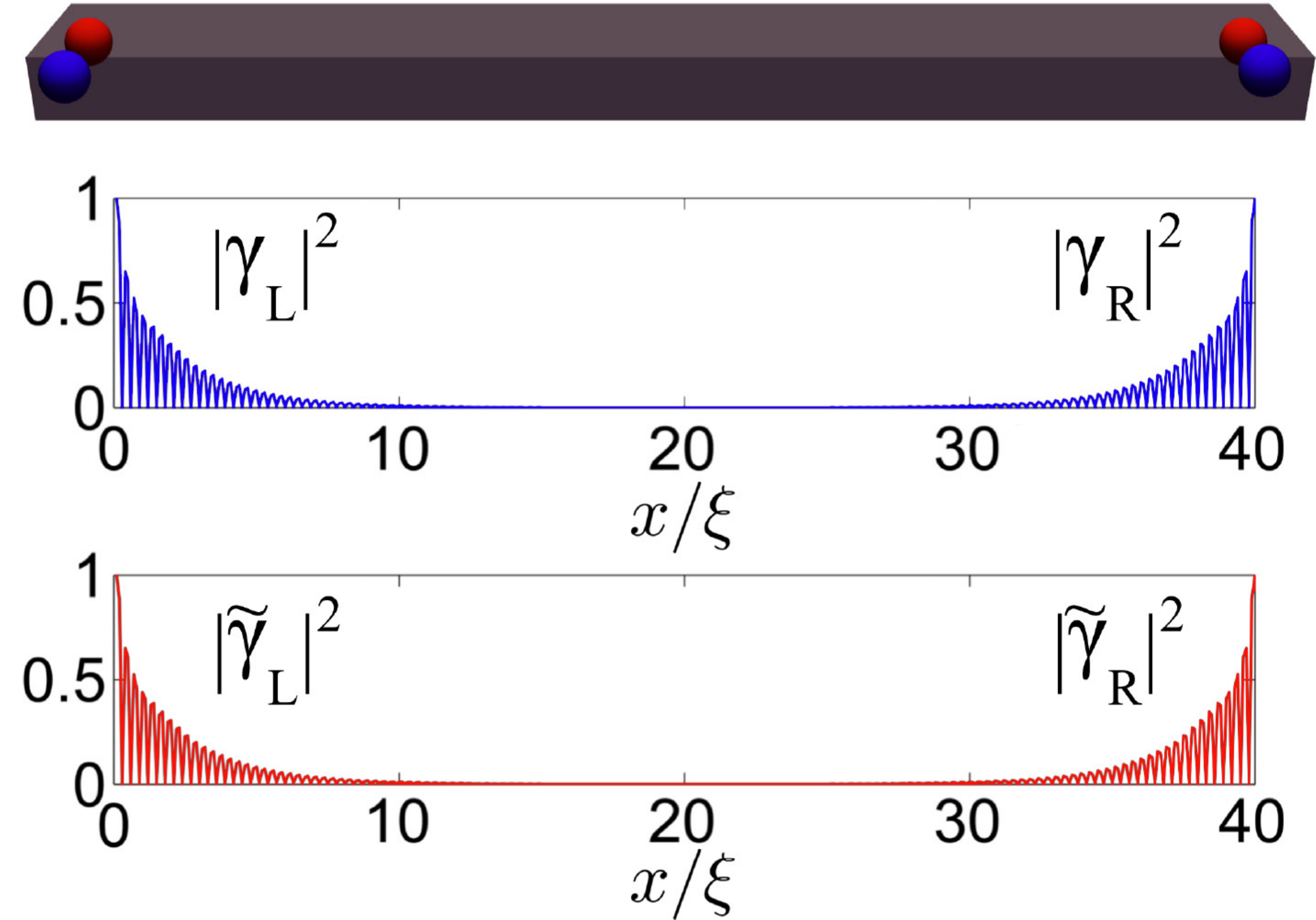}\caption{(a) Two Majorana bound modes exist at each end of the nanowire in the topological regime with $\mu=-2t_w+|\Delta_p^{(0)}|$ as considered in Fig.~\ref{SIspectrum} (b). (b-c) The wave functions of two Majorana modes $\gamma_{L/R}$ and $\tilde\gamma_{L/R}$ at the same end have exactly the same spatial profile, with $\xi$ the coherence length in the nanowire.}
\label{SIboundwavefunction}
\end{figure}

The induced superconductivity in the wire can be obtained by integrating out the degree of freedom of the superconductor substrate. We perform the integration in two steps. First, for the uniform non-centrosymmetric superconductor we can determine its Green's function $G_s(k_x,i_{y0})$ with momentum $k_x$ and at the site $i_{y0}$ below the nanowire by standard recursive method \cite{TD}. Then, the coupling of the nanowire to the superconductor can be reduced to the coupling to the site $i_{y0}$ below the wire and described by the Green's function $G_s(k_x,i_{y0})$. Integrating out the degree of freedom of the sites in the superconductor right below the nanowire yields a self-energy for the Green's function of the nanowire, which gives rise to the proximity effect. The effective Green's function of the nanowire takes the form
\begin{eqnarray}
G_{\rm wire}(i\omega,k_x)=\frac{1}{i\omega -\epsilon_w(k_x) \tau_z - \Sigma(i\omega)},
\end{eqnarray}
where $\epsilon_w=-2t_w\cos k_x-\mu_w$ and the self-energy reads $\Sigma(i\omega) = t_\perp^2 G_s(k_x,i_{y0})$. Finally the spectral function is determined by
\begin {eqnarray}
A(\omega, k_x) =-\frac{1}{2\pi}\Im\{ {\rm Tr} [\tau_z G_{\rm wire}(\omega+i 0^+,k_x)]\},
\end {eqnarray}
with $ \Im$ taking the imaginary part, ${\rm Tr}$ denoting the trace over the spin and Nambu spaces, and $0^+$ a positive infinitesimal. The spectral function determines the bulk band structure, which is numerically shown in Fig.~\ref{SIspectrum} with different chemical potentials of the nanowire. In particular, from the numerical results we find that the nanowire is in the topologically nontrivial regime when $|\mu_w|<2|t_w|$ and $|\Delta_p^{(0)}|>|\Delta_s^{(0)}|$ which leads to the induced pairings in the wire $|\Delta_p|>|\Delta_s|$, while it is in the trivial regime when $|\Delta_s^{(0)}|>|\Delta_p^{(0)}|$ or $|\mu_w|>2|t_w|$ (i.e. the chemical potential is tuned out of the band of the wire). When tuning the chemical potential down to the band bottom, the bulk gap in the nanowire is reduced and closes right at the bottom, implying the critical value of the chemical potential $\mu_w^c=-2t_w$ (similar results can be obtained around $2t_w$, the top of the band) [Fig.~\ref{SIspectrum} (a-c)]. In the topological regime at each end of the nanowire are localized two Majorana zero modes $\gamma_j$ and $\tilde\gamma_j$ ($j=L,R$) which form a Kramers' doublet, with their wave functions shown in Fig.~\ref{SIboundwavefunction}. Further lowering the chemical potential reopens the bulk gap, and the system is driven into a trivial phase [Fig.~\ref{SIspectrum} (d)].

It is interesting that the phase diagram in the nanowire does not depend on parameter details of the couplings between the nanowire and the substrate superconductor, and for $|\Delta_p^{(0)}|>|\Delta_s^{(0)}|$, the topological regime in the nanowire can be obtained in a large parameter range that $-2t_w<\mu_w<2t_w$.
This enables a feasible way to engineer the DIII class topological states in the experiment by tuning $\mu_w$ to be below or above the band bottom of the nanowire.

We note that the time-reversal symmetry is essential for the existence of the Majorana doublets in the topological phase. If time-reversal symmetry is broken, e.g. by introducing a Zeeman term $M_z\sigma_z$, the two Majorana modes at the same end will couple to each other and open a gap. On the other hand, while we consider here the DIII class 1D topological superconductor realized using proximity effect of non-centrosymmetric superconductors, the non-Ablelian statistics predicted in this work are generic results and can be studied with any setup for the 1D TRI topological superconductor as proposed in recent works~\cite{Law2,Kane,Sho,Keselman}.

\section{Non-Abelian statistics}

In this section we show in detail that the Majorana Kramers' doublets obey non-Abelian statistics due to the protection of time-reversal symmetry. In the previous section, we have demonstrated that for the topological phase, at each end of the $Z_2$ Majorana quantum wire are localized two Majorana zero modes $\gamma_j$ and $\tilde\gamma_j$ ($j=1,2$), transformed by time-reversal operator that ${\cal T}^{-1}\gamma_j{\cal T}=\tilde\gamma_j$ and ${\cal T}^{-1}\tilde\gamma_j{\cal T}=-\gamma_j$ \cite{Qi1}. To prove the non-Abelian statistics we first show below a new result that in the DIII class topological superconductor the fermion parity is conserved for each time-reversed sector of the system. With this result we further get that braiding Majorana doublets can generically reduce to two independent processes of exchanging respectively two pairs of Majoranas belonging to two different time-reversed sectors, which leads to the symmetry protected non-Abelian statistics.

\subsection{Fermi parity conservation}

Fermion parity measures the even and odd numbers of the fermions in a quantum system. In a superconductor the fermion number of a ground state can only vary by pairs due to the presence of a pairing gap, which leads to the fermion parity conservation for superconductors. For the DIII class 1D topological superconductor,
we prove here a central result that by grouping all the quasiparticle states into two sectors being time-reversed partners of each other, the fermion parity is conserved for each sector, not only for the total system. It is trivial to know that this result is true if the DIII class topological superconductor is composed of two decoupled copies (e.g. corresponding to spin-up and spin-down, respectively) of 1D chiral $p$-wave superconductors. For the generic case,
the proof is equivalent to showing that in a TRI Majorana quantum wire, the four topological qubit states $|n_1\tilde n_1\rangle$ ($n_1,\tilde n_1=0,1$) are decoupled from each other with the presence of finite TRI perturbations (the change in the fermion parity for each sector necessitates the transition between $|0\tilde1\rangle$ and $|1\tilde0\rangle$ or between $|0\tilde0\rangle$ and $|1\tilde1\rangle$). The coupling Hamiltonian, assumed to depend on a manipulatable parameter $\lambda$, should take the generic TRI form $V(\lambda)=iE_1(\lambda)(\gamma_1\gamma_2-\tilde\gamma_1\tilde\gamma_2)+iE_2(\lambda)(\gamma_1\tilde\gamma_2-\gamma_2\tilde\gamma_1)$,
which splits the two even parity eigenstates $|0\tilde0\rangle$ and $|1\tilde1\rangle$ by an energy $E(\lambda)=2\sqrt{E_1^2+E_2^2}$. Since $|1\tilde0\rangle$ and $|0\tilde1\rangle$ form a Kramers' doublet at arbitrary $\lambda$ value, the transition between them is forbidden by time-reversal symmetry.
Then the fermion parity conservation requires that the following adiabatic condition be satisfied in the
manipulation:
$|\langle1\tilde1|\dot{\lambda}\partial_\lambda|0\tilde0\rangle|\ll2E(\lambda)$,
where $\dot{\lambda}=\partial\lambda/\partial t$.
This is followed by
\begin{eqnarray}\label{eqn:maincondition2}
\tilde R\equiv\frac{1}{2E(\lambda)}\bigr|\frac{\partial\lambda}{\partial t}\frac{\partial\theta}{\partial\lambda}\bigr|\ll1, \ \theta=\tan^{-1}\frac{E_1}{E_2}.
\end{eqnarray}
We show below that the above condition is generically satisfied under realistic conditions.

According to the the previous section, the proximity effect induces $p$-wave and $s$-wave superconducting pairings in the nanowire. The effective tight-binding Hamiltonian of the DIII class Majorana nanowire in the generic case can be written as
\begin{eqnarray}\label{eqn:mainnanowiretightbinding}
H_{\rm wire}^{\rm eff}&=&\sum_{\langle i,j\rangle,\sigma}t_{ij}c_{i\sigma}^\dag c_{j\sigma}+\sum_{\langle i,j\rangle}(t_{ij}^{\rm so}c_{i\uparrow}^\dag c_{j\downarrow}+{\rm H.c.})\nonumber\\
&&+\sum_{\langle i,j\rangle}(\Delta^p_{ij}c_{i\uparrow}c_{j\uparrow}+\Delta^{p*}_{ij}c_{i\downarrow}c_{j\downarrow}+{\rm H.c.})\nonumber\\
&&+\sum_{j}(\Delta_sc_{j\uparrow}c_{j\downarrow}+{\rm H.c.})-\mu\sum_{j,\sigma} n_{j\sigma},
\end{eqnarray}
where the hopping coefficients and the chemical potential are generically renormalized by the proximity effect, with the spin-conserved and the spin-orbit coupled hopping terms satisfying $t_{ij}=t_{ji}=t$ and $t_{ij}^{\rm so}=-t_{ji}^{\rm so}=t_{\rm so}$. For the case with uniform pairing orders, the parameters $\Delta_s$ and $\Delta_p$ can be taken as real. On the other hand, for the present 1D system, one can verify that the phases in the (spin-orbit) hopping coefficients can always be absorbed into electron operators. Therefore, below we consider that all the parameters in $H_{\rm wire}^{\rm eff}$ are real numbers.
Then in terms of the electron operators, the Majorana bound modes take the following general forms
\begin{eqnarray}\label{eqn:main}
\gamma_1&=&\sum_j\bigr[u^{(1)}_{\uparrow}(x_j)c_{\uparrow}(x_j)+u^{(1)}_{\downarrow}(x_j)c_{\downarrow}(x_j)\nonumber\\
&&+u^{(1)}_{\uparrow}(x_j)c^\dag_{\uparrow}(x_j) +u^{(1)}_{\downarrow}(x_j)c^\dag_{\downarrow}(x_j)\bigr],\\
\gamma_2&=&i\sum_j\bigr[u^{(2)}_{\uparrow}(x_j)c_{\uparrow}(x_j)+u^{(2)}_{\downarrow}(x_j)c_{\downarrow}(x_j)\nonumber\\
&&-u^{(2)}_{\uparrow}(x_j)c^\dag_{\uparrow}(x_j) -u^{(2)}_{\downarrow}(x_j)c^\dag_{\downarrow}(x_j)\bigr],
\end{eqnarray}
and $\tilde\gamma_j={\cal T}\gamma_j{\cal T}^{-1}$.
The coupling energies between the Majorana modes at left ($\gamma_{1},\tilde\gamma_1$) and right ($\gamma_2,\tilde\gamma_2$) ends are calculated by $E_1=i\langle\gamma_1|H_{\rm wire}^{\rm eff}|\gamma_2\rangle=-i\langle\tilde\gamma_1|H_{\rm wire}^{\rm eff}|\tilde\gamma_2\rangle$ and $E_2=i\langle\gamma_1|H_{\rm wire}^{\rm eff}|\tilde\gamma_2\rangle=-i\langle\tilde\gamma_1|H_{\rm wire}^{\rm eff}|\gamma_2\rangle$.

\begin{figure}[ht]
\includegraphics[width=0.9\columnwidth]{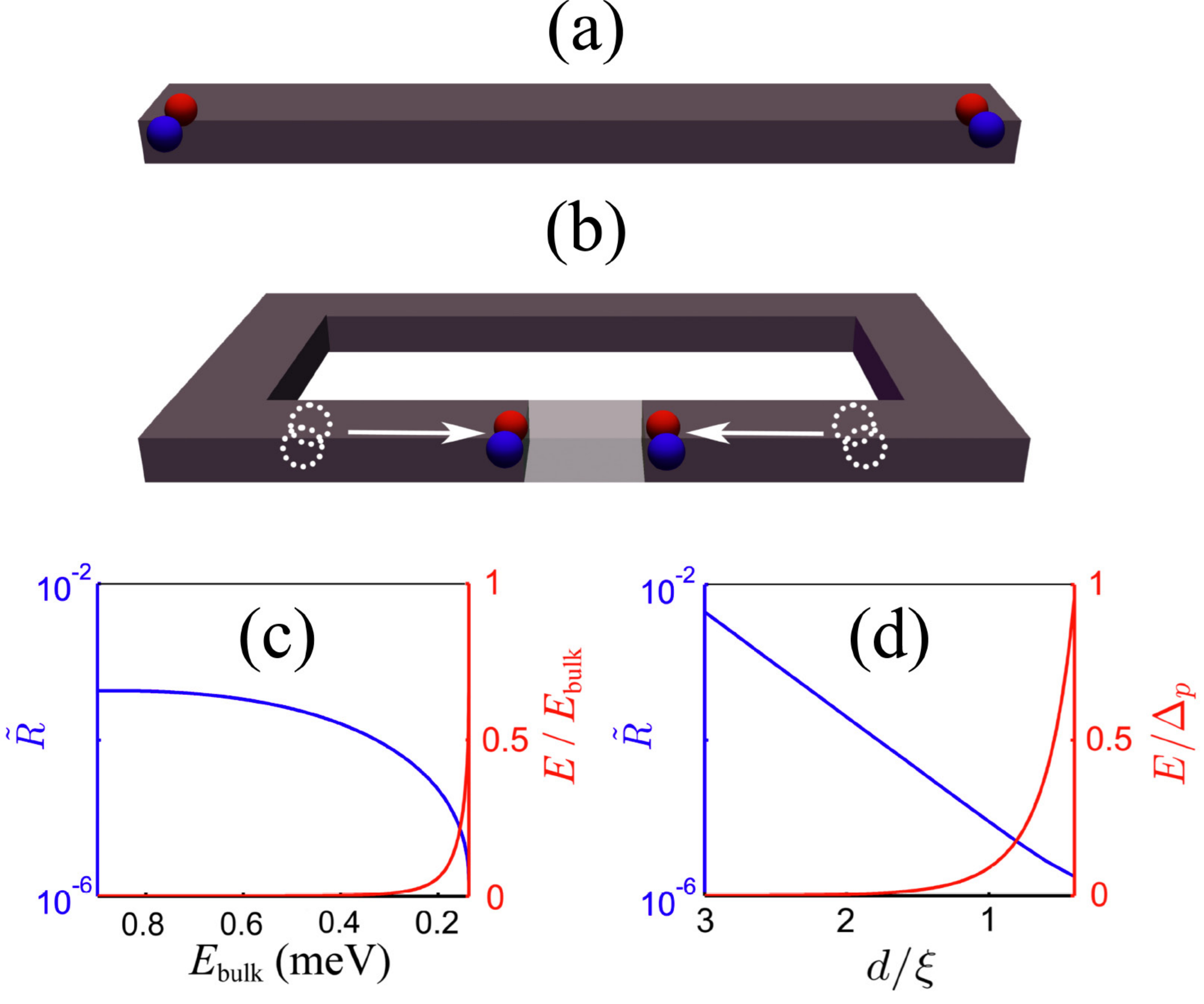}\caption{Adiabatic condition and fermion parity conservation for each sector of time-reversal partners. (a-b) The couplings $E_{1,2}$ between Majorana end modes are manipulated by tuning the chemical which changes the bulk gap ($\lambda=E_{\rm bulk}$) in the nanowire (a), and by varying the length $(\lambda=d)$ of a trivial region (gray color) which separates the two pairs of Majorana modes ({b}). (c-d) The energy splitting $E$ between $|0\tilde0\rangle$ and $|1\tilde1\rangle$ (red curves) and the ratio $\tilde R$ (blue curves), as functions of $E_{\rm bulk}$ ({c}), and versus the trivial region distance $d$ ({d}). The parameters in the nanowire are taken that the proximity induced $p$-wave pairing $\Delta_p=1.0$meV, $s$-wave pairing $\Delta_s=0.5$meV, and the spin-orbit coupling energy $E_{\rm so}=0.1$meV. In the numerical simulation we assume that the coupling energy $E$ is tuned from $0$ to $1.0$meV in the time $1.0\mu$s. We also numerically confirmed the adiabatic condition $\tilde R\ll1$ with other different parameter regimes.}
\label{Parity}
\end{figure}
It can be found that the coefficients $E_{1,2}$ are proportional to the overlapping integrals of the left- and right-end Majorana wave functions, which decay exponentially with the distance $d$ between the Majorana modes.
Since $\gamma_j$ and $\tilde\gamma_j$ are connected by ${\cal T}$-transformation, their wave functions have exactly the same spatial profile, which leads to the same exponential form of the coefficients $E_{1,2}(\lambda)=\alpha_{1,2}(\lambda)e^{-d/\xi}$ with $\xi$ the coherence length in the nanowire. The pre-factors $\alpha_j(\lambda)$ depend on the local couplings, i.e. the hopping terms and pairings in $H_{\rm wire}^{\rm eff}$, between electrons belonging to the same (for $j=1$) or different (for $j=2$) sectors of the time-reversal partners. For the realistic conditions, we consider that the chemical potential in the nanowire is far below the half-filling condition and thus the Fermi momentum satisfying $k_Fa\ll1$, and the coherence length (in the order of $1.0\mu$m) is typically much larger than the lattice constant $\xi\gg a$ ($a\sim0.5$nm). Under these conditions we can verify that to $E_{1,2}$ the contributions of the spin-orbit coupling and $p$-wave pairing terms in $H_{\rm wire}^{\rm eff}$ vanish, and we find (details can be found in the Appendix section)
\begin{eqnarray}
E_1&\simeq&t\sum_{\langle i,j\rangle\sigma}u^{(1)}_{\sigma}(x_i)u^{(2)}_{\sigma}(x_j),\label{eqn:maincoupling1}\\
E_2&\simeq&\Delta_s\sum_{\langle i,j\rangle\sigma}u^{(1)}_{\sigma}(x_i)u^{(2)}_{\sigma}(x_j).\label{eqn:maincoupling2}
\end{eqnarray}
Therefore while the magnitudes of $E_{1,2}$ can vary with $d$ and the bulk gap, their ratio $E_1/E_2$ is nearly a constant, and we always have $\partial_\lambda\theta\approx0$, which validates the adiabatic condition. The above results are consistent with the fact that when $\Delta_s=0$ the original Hamiltonian~\eqref{eqn:mainnanowiretightbinding} can be block diagonalized and then $E_2\equiv0$. The adiabatic condition is clearly confirmed with the numerical results in Fig.~\ref{Parity}. The fermion parity conservation for each sector shows that an isolated DIII class 1D Majorana wire should stay in one of the four fermion parity eigenstates germinated by non-local complex fermion operators $f_j$ and $\tilde f_j$, given that time-reversal symmetry is not broken. In particular, one can always prepare a nanowire initially in the ground state $|0\tilde0\rangle$ or $|1\tilde1\rangle$ by controlling the initial couplings $E_{1,2}(\lambda)$, and then manipulate the states adiabatically. A weak time-reversal breaking term, e.g. induced by a stray field if existing in the environment, may induce couplings between qubit states with the same total fermion parity. For typical semiconductor nanowires, e.g. the InSb wire which has a large Lande factor $g\approx50$~\cite{Kouwenhoven}, one can verify that the time-reversal breaking couplings are negligible if the field strength is much less than $0.01$T. For other types of nanowires with smaller $g$-factors, the couplings are not harmful with even larger stray fields.

It is worthwhile to note that in the above discussion we did not consider the quasiparticle poisoning which may change fermion parity and lead to decoherence of Majorana qubit states. At low temperature, the dominant effect in the quasiparticle poisoning comes from the single electron tunneling between the nanowire and the substrate superconductor~\cite{Rainis}. The decoherence time in the chiral Majorana nanowires ranges from $10$ns to $0.1$ms, depending on parameter details~\cite{Rainis}. For the DIII class nanowires, without suppression of external magnetic field, the proximity induced gap in the similar parameter regime is expected to be larger compared with that in chiral nanowires, which suggests a longer decoherence time in the DIII class Majorana nanowires~ \cite{Law2,Kane,Sho,Keselman}.

To ensure that the decoherence effect induced by quasiparticle poisoning does not lead to serious problems, one requires that the adiabatic manipulation time for Majorana modes should be much less than the decoherence time. For the DIII class Majorana nanowires, the adiabatic time depends on the two characteristic time scales. One is determined by the bulk gap $\tau_1^{\rm ad}=h/E_g$, and another $\tau_2^{\rm ad}$ corresponds to the fermion parity conservation for each time-reversal sector. The typical time scale $\tau_1^{\rm ad}$ in the DIII class Majorana nanowires can be about $0.1$ns and is much less than the decoherence time. Furthermore, if using the parameter regime in Fig.~\ref{Parity}, one can estimate that $\tau_2^{\rm ad}<1.0$ns. On the other hand, for the proposals considered in Refs.~\cite{Law2,Kane,Keselman}, the effective Hamiltonian has no $s$-wave pairing order, and the time scale $\tau_2^{\rm ad}$ indeed renders the magnitude of $\tau_1^{\rm ad}$. These estimates imply that the adiabatic manipulation of Majorana modes may be reached in DIII class 1D topological superconductors.

\subsection{Braiding statistics}

\begin{figure}[ht]
\includegraphics[width=0.95\columnwidth]{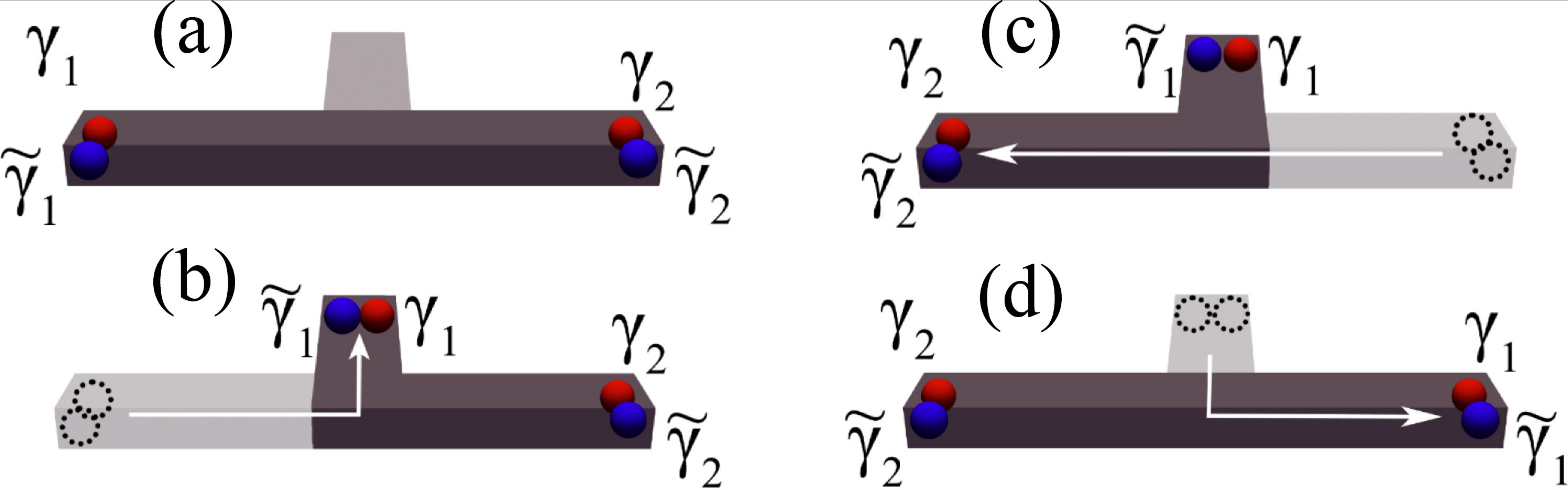}\caption{(a-d) Braiding Majorana end modes through gating a T-junction following the study by Alicea et al.~\cite{Alicea1}. The dark (light gray) area of the nanowires depicts the topological (trivial) region, which can be controlled by tuning the chemical potential in the nanowire. The arrows depict the direction that the Majorana fermions are transported to in the braiding process.}
\label{braiding}
\end{figure}

Note that braiding Majorana end modes is not well-defined for a single 1D nanowire and, as first recognized by Alicea et al., the minimum setup for braiding requires a trijunction, e.g. a T-junction composed of two nanowire segments \cite{Alicea1}. The braiding can be performed by transporting the Majorana zero modes following the steps as illustrated in Fig.~\ref{braiding}(a-d).

The fermion parity conservation for each sector shown above implies that the exchange of Majorana end modes in DIII class topological superconductor generically reduces to two independent processes of braiding Majoranas of two different sectors, respectively. This is because, first of all, braiding adiabatically the Majorana pairs, e.g. $\gamma_1,\tilde\gamma_1$ and $\gamma_2,\tilde\gamma_2$ in Fig.~\ref{braiding}, does not affect the bulk states which are gapped. Furthermore, assuming that other Majorana modes are located far away from $\gamma_{1,2}$ and $\tilde\gamma_{1,2}$, the braiding evolves only the Majoranas which are exchanged. Finally, due to the fermion parity conservation, in the braiding the fermion modes $f_1$ and $\tilde f_1$ are decoupled and their dynamics can be derived independently. By a detailed derivative we show that after braiding the topological qubit states evolve according to (see the Supplementary Material~\cite{SI})
\begin{equation}\label{eq:braidingphase}
  \begin{split}
|1\rangle|\tilde1\rangle_{\rm final}&=|1\rangle|1\rangle_{\rm initial},\\
|1\rangle|\tilde0\rangle_{\rm final}&=i|1\rangle|\tilde0\rangle_{\rm initial},\\
|0\rangle|\tilde1\rangle_{\rm final}&=-i|0\rangle|\tilde1\rangle_{\rm initial},\\
|0\rangle|\tilde0\rangle_{\rm final}&=|0\rangle|\tilde0\rangle_{\rm initial}.
\end{split}
\end{equation}
We therefore obtain the braiding matrix by $U_{12}(T,\tilde T)=\exp(\frac{\pi}{4}\gamma_1\gamma_2)\exp(\frac{\pi}{4}\tilde\gamma_1\tilde\gamma_2)$, which is time-reversal invariant. Note that the oppositely-handed braiding process of $U_{1,2}$ is given by $U^\dag_{12}(T,\tilde T)=\exp(-\frac{\pi}{4}\gamma_1\gamma_2)\exp(-\frac{\pi}{4}\tilde\gamma_1\tilde\gamma_2)$, which describes a process that one first transports $\gamma_2$ and $\tilde\gamma_2$ to the end of the vertical wire, then transports the two modes $\gamma_1$ and $\tilde\gamma_1$ to the right hand end, and finally, the two modes $\gamma_2$ and $\tilde\gamma_2$ are transported to the left hand end of the horizontal wire. The braiding matrix $U_{12}(T,\tilde T)$ exactly reflects that the two pairs of Majoranas $\gamma_1,\tilde\gamma_1$ and $\gamma_2,\tilde\gamma_2$ are braided independently. Actually, this braiding rule can be visualized most straightforwardly if we consider the simplest situation that the DIII class topological superconductor is composed of two decoupled copies of 1D chiral $p$-wave superconductors. In this case the whole braiding must be a product of two independent processes of braiding $\gamma_{1,2}$ and $\tilde\gamma_{1,2}$, respectively, yielding the above braiding matrix following the studies in the Refs.~\cite{Ivanov,Alicea1}. This braiding is nontrivial and leads to the symmetry protected non-Abelian statistics as presented below.

\begin{figure}[ht]
\includegraphics[width=0.8\columnwidth]{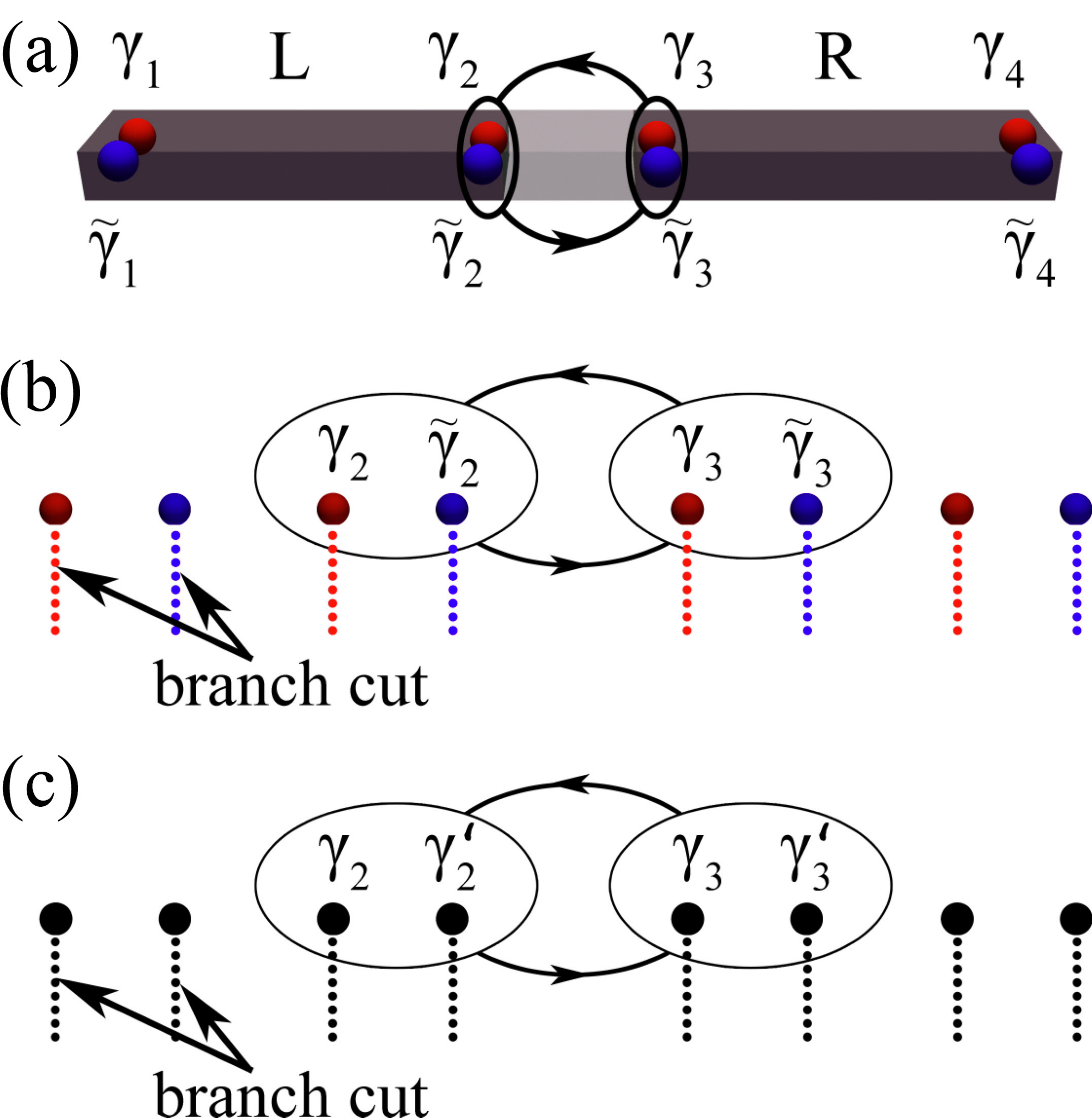}\caption{Non-Abelian statistics in DIII class 1D topological superconductor. (a) Majorana end modes $\gamma_2,\tilde\gamma_2$ and $\gamma_3,\tilde\gamma_3$ are braided through similar processes shown in Fig.~\ref{braiding}. (b) Braiding Majorana modes in DIII class superconductor is equivalent to two independent processes of exchanging $\gamma_2,\gamma_3$ and $\tilde\gamma_2,\tilde\gamma_3$, respectively. In the depicted process $\gamma_2$ ($\tilde\gamma_2$) crosses only the branch cut of $\gamma_3$ ($\tilde\gamma_3$), and therefore acquires a minus sign after braiding. (c) In contrast, if braiding two Majorana pairs in a chiral superconductor, for the depicted process $\gamma_{2}$ ($\gamma_2'$) crosses the branch cuts of both $\gamma_3$ and $\gamma_3'$, and then no sign change occurs for the Majorana operators after braiding \cite{Ivanov}. Therefore, braiding twice two Majorana pairs always returns to the original state.}
\label{nonAbelian}
\end{figure}

We consider two DIII class wires with eight Majorana modes $\gamma_{1,...,4}$ and $\tilde\gamma_{1,...,4}$ [Fig.~\ref{nonAbelian}(a)], which define four complex fermion modes by $f_1=\frac{1}{2}(\gamma_1+i\gamma_2), f_2=\frac{1}{2}(\gamma_3+i\gamma_4)$, and $\tilde f_{1,2}={\cal T}^{-1}f_{1,2}{\cal T}$.
The Hilbert space of the four complex fermions is spanned by sixteen qubit states $|n_1\tilde n_1\rangle_L|n_2\tilde n_2\rangle_R$ ($n_{1,2},\tilde n_{1,2}=0,1$), where $L/R$ represents the left/right nanowire segment. If the initial state of the system is $|0\tilde0\rangle_L|0\tilde0\rangle_R$, for instance, by braiding the two pairs of Majoranas $\gamma_2,\tilde\gamma_2$ and $\gamma_3,\tilde\gamma_3$ we get straightforwardly
\begin{eqnarray}\label{eqn:mainbraid1}
U_{23}(T,\tilde T)|0\tilde0\rangle_L|0\tilde0\rangle_R&=&\frac{1}{2}\bigr(|0\tilde0\rangle_L|0\tilde0\rangle_R+|1\tilde1\rangle_L|1\tilde1\rangle_R\nonumber\\ &+&i|1\tilde0\rangle_L|1\tilde0\rangle_R-i|0\tilde1\rangle_L|0\tilde1\rangle_R\bigr).
\end{eqnarray}
It is interesting that the above state is generically a four-particle entangled state, which shows the natural advantage in generating multi-particle entangled state using DIII class topological superconductors. Furthermore, a full braiding, i.e. braiding twice $\gamma_2,\tilde\gamma_2$ and $\gamma_3,\tilde\gamma_3$ yields the final state $|1\tilde1\rangle_L|1\tilde1\rangle_R$, which distinguishes from the initial state in that each copy of the $p$-wave superconductor changes fermion parity. After braiding four times the two pairs of Majoranas the ground state returns to the original state. On the other hand, it is also straightforward to verify that $U_{12}U_{23}\neq U_{23}U_{12}$, implying the non-commutability of the braiding processes. These results demonstrate the non-Abelian statistics obeyed by Majorana doublets.

From the above discussion we find that in the braiding the Majorana modes $\gamma_j$ are unaffected by their time-reversal partners $\tilde \gamma_j$, which is an essential difference from the situation in exchanging two pairs of Majoranas in a chiral superconductor, and makes the braiding operator in the TRI topological superconductor nontrivial. This property can be pictorialized by assigning branch cuts for the Majorana modes braided through the junction~\cite{Alicea1}, as illustrated in Fig.~\ref{nonAbelian}(b-c). When exchanging Majorana modes $\gamma_2,\gamma_3$ and $\tilde\gamma_2,\tilde\gamma_3$ in the DIII class superconductor, $\gamma_2$ ($\tilde\gamma_2$) crosses only the branch cut of $\gamma_3$ ($\tilde\gamma_3$) and therefore acquires a minus sign after braiding. In contrast, if braiding two Majorana pairs in a chiral superconductor, for the process in Fig.~\ref{nonAbelian}(c) $\gamma_{2}$ ($\gamma_2'$) crosses the branch cuts of both $\gamma_3$ and $\gamma_3'$, and then no sign change occurs for the Majorana operators after braiding \cite{Ivanov}. Therefore, a full braiding of two Majorana pairs always returns to the original state.

It is worthwhile to note that to realize a DIII class superconductor applies no external magnetic field, which might be advantageous to construct realistic Majorana network to implement braiding operations. In comparison, for the chiral topological superconductor observed in a spin-orbit coupled semiconductor nanowire using $s$-wave superconducting proximity effect \cite{Kouwenhoven,Deng,Das}, the external magnetic field should be applied perpendicular to the spin quantization axis by spin-orbit interaction, driving optimally the nanowire into topological phase \cite{Kouwenhoven,Das,Liu}. It is shown that for a network formed by multiple nanowire segments, such optimal condition cannot be reached for all segments without inducing detrimental orbital effects, which creates further experimental challenges in braiding Majoranas \cite{Liu}. It is clear that such intrinsic difficulty is absent in the present DIII class TRI topological superconductor, and one may have more flexibility in constructing 2D and even 3D Majorana networks for topological quantum computation.

\section{Josephson effect in DIII class topological superconductor}

It is important to study how to detect the topological qubit states in a DIII class Majorana quantum wire. The ground states of a single DIII class Majorana quantum wire include two even ($|0\tilde0\rangle$ and $|1\tilde1\rangle$) and two odd ($|0\tilde1\rangle$ and $|1\tilde0\rangle$) parity eigenstates. In a chiral topological superconductor the states of the same fermion parity are not distinguishable. On other hand, in the generic case the two different time-reversal sectors do not correspond to different measurable good quantum numbers (e.g. spin). Therefore, the two qubit states with same total fermion parity, e.g. $|0\tilde1\rangle$ and $|1\tilde0\rangle$, cannot be distinguished via direct quantum number measurements. However, according to the fermion parity conservation shown in section III(A), in a 1D TRI topological superconductor the two even/odd parity states are decoupled due to time-reversal symmetry, implying that such two states should be distinguishable. We show in this section that all the four topological qubit states can be measured by the Josephson effect in DIII class topological superconductors.

We consider a Josephson junction illustrated in Fig.~\ref{Josephson} (a) formed by DIII class superconductor. As derived in the Appendix section, the effective coupling Hamiltonian of the Josephson junction is given by
\begin{figure}[ht]
\includegraphics[width=0.9\columnwidth]{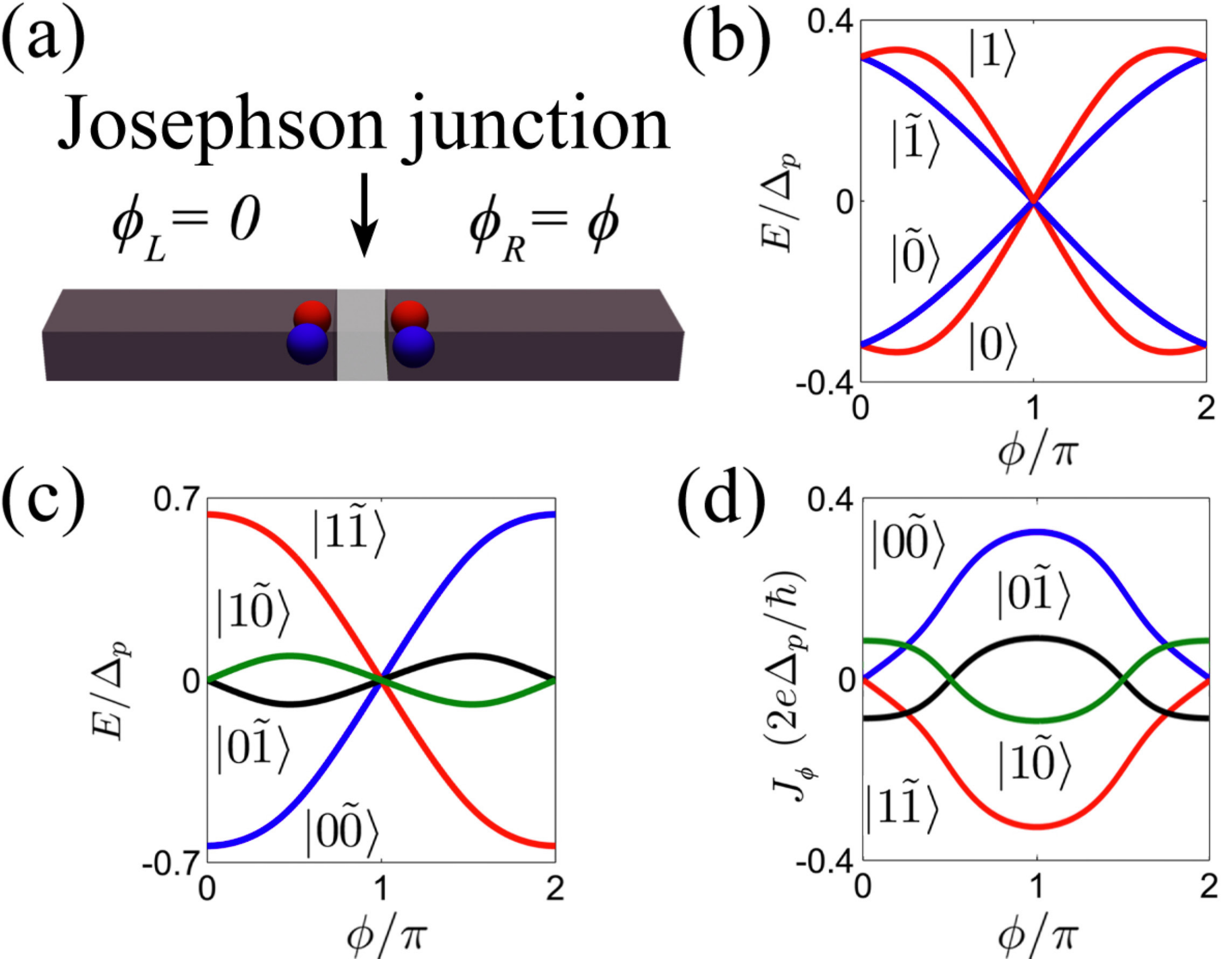}\caption{Josephson measurement of the topological qubit states in DIII class 1D topological superconductor. (a) The sketch of a Josephson junction with phase difference $\phi$. (b) The single particle Andreev bound state spectra versus the phase difference $\phi$. (c) The energy spectra of the four qubit states $|n_1\tilde n_1\rangle$ ($n_1,\tilde n_1=0,1$) according to the results in (b). (d) The Josephson currents (in units of $2e\Delta_p/\hbar$) for different topological qubit states. Parameters used in the numerical calculation are taken that $\Delta_p=1.0$meV, $\Delta_s=0.25$meV, $E_{\rm so}=0.1$meV, the width of the junction $d=0.5\xi$, and in middle trivial region (gray color) of the junction the chemical potential is set to be at the band bottom.}
\label{Josephson}
\end{figure}
\begin{eqnarray}\label{eqn:mainJosephson1}
H_{\rm eff}(\phi)&=&i\Gamma_0\cos\frac{\phi}{2}(\gamma_L\gamma_R-\tilde\gamma_L\tilde\gamma_R)\nonumber\\
&&+i\Gamma_1\sin\phi(\gamma_L\tilde\gamma_L-\gamma_R\tilde\gamma_R),
\end{eqnarray}
where $\phi$ is the phase difference across the junction, and $L/R$ represents the left/right hand lead of the junction. The $\Gamma_0$-term in $H_{\rm eff}$ represents the first-order direct coupling between Majorana fermions at different junction leads. It can be seen that the direct coupling term is of $4\pi$ periodicity, which can be understood in the following way. When the phase difference across the junction advances $2\pi$, the Cooper pair wave function changes $2\pi$ across the junction, while for single electron operators the phase varies only $\pi$. This implies that the coupling coefficients also change $\pi$ phase and thus reverse sign, leading to the $4\pi$ periodicity of the direct coupling term. The $\Gamma_1$-term is resulted from the second-order perturbation of the tunneling process, and this term vanishes if the $s$-wave pairing $\Delta_s=0$. This is because, the couplings such as $i\gamma_j\tilde\gamma_j$ ($j=L,R$) breaks time-reversal symmetry,
while the direct coupling between $\gamma_j$ and $\tilde\gamma_j$ does not experience the phase difference across the junction and should preserve
time-reversal symmetry. Actually, a uniform pairing phase in one end of the junction can be removed by a constant gauge transformation. Therefore the coupling between Majorana fermions at the same end can only be induced by electron tunneling and the minimum requirement is to consider the second-order tunneling process. In the second-order perturbation $\gamma_L$ and $\tilde\gamma_L$ ($\gamma_R$ and $\tilde\gamma_R$) couple to electron modes $c_R$ and $\tilde c_R$ in the right hand end ($c_L$ and $\tilde c_L$ in the left hand end), respectively. When a nonzero $s$-wave pairing is present in the nanowires, the electrons $c_{L/R}$ and $\tilde c_{L/R}$ form a Cooper pair and condense. This process leads to the effective coupling between Majorana zero modes localized at the same end, with the coupling strength proportional to $s$-wave order parameter. Finally, note that the system restores time-reversal symmetry at $\phi=m\pi$, which explains why the $\Gamma_1$-term is proportional to $\sin\phi$, and has $2\pi$ periodicity. All these properties have been confirmed with numerical results.

Redefining the Majorana bases by $\gamma'_1=\gamma_L+\tilde\gamma_R$ and $\gamma'_2=\gamma_R+\tilde\gamma_L$, we recast the above Hamiltonian into $H_{\rm eff}=i(\Gamma_0\cos\phi/2+\Gamma_1\sin\phi)\gamma'_1\gamma'_2-i(\Gamma_0\cos\phi/2-\Gamma_1\sin\phi)\tilde\gamma'_1\tilde\gamma'_2$.
The Andreev bound state spectra are obtained straightforwardly by
\begin{eqnarray}
E_{n_{f'},\tilde n_{f'}}(\phi)&=&\bigr(\Gamma_0\cos\frac{\phi}{2}+\Gamma_1\sin\phi\bigr)(2n_{f'}-1)\nonumber\\
&+&\bigr(\Gamma_0\cos\frac{\phi}{2}-\Gamma_1\sin\phi\bigr)(2n_{\tilde f'}-1),
\end{eqnarray}
which is shown numerically in Fig.~\ref{Josephson} (b-c). Here $n_{f',\tilde f'}$ are complex fermion number operators for $f'$ and $\tilde f'$ modes, respectively. The Josephson currents are obtained by the slope of the Andreev bound state spectra. In particular, we have that the Josephson currents $J^{\rm even}_\phi=\pm\frac{e}{\hbar}\Gamma_0\sin\frac{\phi}{2}$ for the even parity states $|0\tilde0\rangle$ and $|1\tilde1\rangle$, and $J^{\rm odd}_\phi=\pm\frac{e}{\hbar}\Gamma_1\cos\phi$ for the odd parity states $|0\tilde1\rangle$ and $|1\tilde0\rangle$, respectively [Fig.~\ref{Josephson}(d)].

It is remarkable that the currents for odd parity states are of $2\pi$ periodicity, half of those for even parity states [Fig.~\ref{Josephson}(d)]. This reflects that $J^{\rm even}_\phi$ is contributed from the direct Majorana coupling induced by first-order single-electron tunneling~\cite{Kitaev1}, while $J^{\rm odd}_\phi$ is a consequence of the second-order tunneling process which corresponds to the Cooper pair tunneling. This nontrivial property is essentially different from the the Josephson physics with multiple Majorana end modes studied by D. Sticlet et al. in the BDI class Majorana chains~\cite{Sticlet}, where a multi-copy version of the fractional Josephson effect with $4\pi$ periodicity is investigated. The reason is because in a BDI class topological superconductor the time-reversal symmetry operator ${\cal T}^2=1$ and the different copies of the superconductor are not related by time-reversal symmetry (nor by any other symmetry), while in the DIII class topological superconductor the two copies are related by ${\cal T}$-symmetry. The present result is also consistent with the fact that the time-reversal symmetry is restored with $|0\tilde1\rangle$ and $|1\tilde0\rangle$ forming Kramers' doublet at $\phi=m\pi$, which necessitates the $2\pi$ periodicity in their spectra. Furthermore, the two qubit states with the same total parity (e.g. $|0\tilde0\rangle$ and $1\tilde1\rangle$) are distinguished by the direction of the currents. The qualitative difference in the Josephson currents imply that the four topological qubit states can be measured in the experiment. \\
\\

\section{Conclusions}

In summary, we have shown that Majorana doublets obtained in the DIII class 1D topological superconductors obey non-Abelian statistics, due to the protection of time-reversal symmetry. The key results are that the fermion parity is conserved for each copy of the $Z_2$ TRI topological superconductor, and the exchange of Majorana end modes can generically reduce to two independent processes of braiding Majoranas of two different copies, respectively. These results lead to the symmetry protected non-Abelian statistics for the Majorana doublets, and the braiding statistics are protected by time-reversal symmetry. Furthermore, we unveiled an intriguing phenomenon in the Josephson effect, that the periodicity of Josephson currents depends on the fermion parity of the 1D TRI topological superconductors. We found that this effect can provide direct measurements of the topological qubit states in the DIII class Majorana quantum wires. Our results will motivate further studies in both theory and experiments on the braiding statistics and nontrivial Josephson effects in the wide classes of symmetry-protected topological superconductors.\\
\\

\section*{ACKNOWLEDGEMENT}

We appreciate the very helpful discussions with P. A. Lee, L. Fu, Z. -X. Liu, A. Potter, Z. -C. Gu, M. Cheng, C. Wang and X. G. Wen. The authors thank the support of HKRGC through DAG12SC01, Grant 605512, and HKUST3/CRF09.


\noindent

\onecolumngrid

\renewcommand{\thesection}{A-\arabic{section}}
\setcounter{section}{0}  
\renewcommand{\theequation}{A\arabic{equation}}
\setcounter{equation}{0}  
\renewcommand{\thefigure}{A\arabic{figure}}
\setcounter{figure}{0}  

\indent

\section*{\Large\bf Appendix}

In the Appendix section we provide the details of showing the fermion parity conservation for each sector of the time-reversal partners, and deriving the Josephson effect for DIII class 1D topological superconductors.

\section{Fermi parity conservation}

We consider a single Majorana quantum wire, which hosts four Majorana end modes denoted by $\gamma_{1,2}$ and $\tilde\gamma_{1,2}$, and transformed via ${\cal T}^{-1}\gamma_j{\cal T}=\tilde\gamma_j$ and ${\cal T}^{-1}\tilde\gamma_j{\cal T}=-\gamma_j$.
With the four Majorana states we can define two non-local complex fermions by $f_1=\frac{1}{2}(\gamma_1+i\gamma_2), \tilde f_1=\frac{1}{2}(\tilde\gamma_1-i\tilde\gamma_2)$, which germinate four topological qubit states $|n_1\tilde n_1\rangle$ with $n_1,\tilde n_1=0,1$.
The proof of fermion parity conservation for each sector is equivalent to showing that the four topological qubit states $|n_1\tilde n_1\rangle$ are generically decoupled from each other in the presence of TRI perturbations.

Note that the coupling between the Majorana modes localized at the same end of the nanowire, $\gamma_j$ and $\tilde\gamma_j$, breaks time-reversal symmetry. The coupling Hamiltonian in terms of Majorana end modes should take the following generic TRI form
\begin{eqnarray}\label{eqn:}
V(\lambda)=iE_1(\lambda)(\gamma_1\gamma_2-\tilde\gamma_1\tilde\gamma_2)+iE_2(\lambda)(\gamma_1\tilde\gamma_2-\gamma_2\tilde\gamma_1),
\end{eqnarray}
where we assume that the couplings coefficients $E_{1,2}(\lambda)$ depend on an experimentally manipulatable parameter $\lambda$ (e.g. the bulk gap in the nanowire or the distance between the Majorana modes).
The above Hamiltonian can be rewritten in the block diagonal form with new Majorana bases that
\begin{eqnarray}\label{eqn:}
V(\lambda)=iE(\lambda)\bigr[\gamma^{(1)}\gamma^{(2)}-\tilde\gamma^{(1)}\tilde\gamma^{(2)}\bigr]
\end{eqnarray}
where $\gamma^{(1)}=\gamma_1, \tilde\gamma^{(1)}=\tilde\gamma_1, \gamma^{(2)}=\sin\theta\gamma_2+\cos\theta\tilde\gamma_2,
\tilde\gamma^{(2)}=\sin\theta\tilde\gamma_2-\cos\theta\gamma_2$, and $E=\sqrt{E_1^2+E_2^2}$. The mixing angle $\theta$ is defined via $\tan\theta=E_1/E_2$.
The complex fermions $f^{(1)}$ and $\tilde f^{(1)}$ in the eigen-basis are then defined by
\begin{eqnarray}\label{eqn:}
f^{(1)}=\frac{1}{2}\bigr[\gamma^{(1)}+i\gamma^{(2)}\bigr], \ \tilde f^{(1)}=\frac{1}{2}\bigr[\tilde\gamma^{(1)}-i\tilde\gamma^{(2)}\bigr].
\end{eqnarray}
It is easy to know that the even parity eigenstates $|0\tilde0\rangle$ and $|1\tilde1\rangle$ germinated by $f^{(1)}$ and $\tilde f^{(1)}$ acquire an energy splitting $2E(\lambda)$, while the odd parity states $|0\tilde1\rangle$ and $|1\tilde0\rangle$ are still degenerate due to time-reversal symmetry. To prove the fermion parity conservation for each sector, we need to confirm that all the four topological qubit states $|n_1\tilde n_1\rangle$ can evolve adiabatically when the coupling Hamiltonian $V(\lambda)$ varies with the parameter $\lambda$. Since $|1\tilde0\rangle$ and $|0\tilde1\rangle$ form a Kramers' doublet, the transition between them is forbidden by the time-reversal symmetry. Therefore, we only need to consider the adiabatic condition for the two even parity states. The fermion parity conservation for each sector is guaranteed when the following adiabatic condition is satisfied in the manipulation
\begin{eqnarray}\label{eqn:condition1}
\bigr|\langle1\tilde1|\frac{\partial\lambda}{\partial t}\frac{\partial}{\partial\lambda}|0\tilde0\rangle\bigr|\ll2|E(\lambda)|.
\end{eqnarray}
It should be noted that the adiabatic condition needs to be justified only in the presence of finite couplings. When $E(\lambda)\rightarrow0$, the couplings between Majorana end modes vanish and then all the topological qubit states are automatically decoupled from each other. One can verify that
\begin{eqnarray}\label{eqn:}
\frac{\partial f^{(1)}}{\partial\lambda}&=&\frac{i}{2}\frac{\partial\theta}{\partial\lambda}(\cos\theta\gamma_2-\sin\theta\tilde\gamma_2) =\frac{1}{2}\frac{\partial\theta}{\partial\lambda}(\tilde f^{(1)\dag}-\tilde f^{(1)}),\\
\frac{\partial\tilde f^{(1)}}{\partial\lambda}&=&-\frac{i}{2}\frac{\partial\theta}{\partial\lambda}(\cos\theta\tilde\gamma_2+\sin\theta\gamma_2) =-\frac{1}{2}\frac{\partial\theta}{\partial\lambda}(f^{(1)\dag}-f^{(1)}).
\end{eqnarray}
With some calculation one can show that in the above formulas the derivatives of the bases $\gamma_j,\tilde\gamma_j$ with respect to $\lambda$ will not contribute to the left hand side of Eq.~\eqref{eqn:condition1}, and therefore are neglected. The condition \eqref{eqn:condition1} then reads
\begin{eqnarray}\label{eqn:condition2}
\bigr|\frac{\partial\lambda}{\partial t}\frac{\partial\theta}{\partial\lambda}\bigr|\ll4|E(\lambda)|.
\end{eqnarray}
We show below that the above condition is generically satisfied in the realistic materials.

With the proximity induced $p$-wave and $s$-wave superconducting pairings, the effective tight-binding Hamiltonian in the nanowire can be generically written as
\begin{eqnarray}\label{eqn:nanowiretightbinding}
H_{\rm wire}^{\rm eff}&=&\sum_{\langle i,j\rangle,\sigma}t_{ij}c_{i\sigma}^\dag c_{j\sigma}+\sum_{\langle i,j\rangle}(t_{ij}^{\rm so}c_{i\uparrow}^\dag c_{j\downarrow}+{\rm H.c.})+\sum_{\langle i,j\rangle}(\Delta^p_{ij}c_{i\uparrow}c_{j\uparrow}+\Delta^{p*}_{ij}c_{i\downarrow}c_{j\downarrow}+{\rm H.c.})+\sum_{j}(\Delta_sc_{j\uparrow}c_{j\downarrow}+{\rm H.c.})\nonumber\\
&&-\mu\sum_{j,\sigma} n_{j\sigma}+\sum_{j,\sigma}V_{j}^{\rm dis}n_{j\sigma},
\end{eqnarray}
where the hopping coefficients and the chemical potential are generically renormalized by the proximity effect. Without loss of generality, in the above Hamiltonian we have taken into account the spin-orbit interaction described by the $t_{ij}^{\rm so}$ term, and the random on-site disorder potential $V_{j}^{\rm dis}$ with $\langle V_j^{\rm dis}\rangle=0$. For the case with uniform pairing orders, the parameters $\Delta_s$ and $\Delta_p$ can be taken as real. On the other hand, for the present 1D system, one can verify that the phases in the (spin-orbit) hopping coefficients can always be absorbed into electron operators. Therefore, in the following study we consider that all the parameters in $H_{\rm wire}^{\rm eff}$ are real numbers.

In the topological regime, at each end of the wire we obtain two Majorana zero modes which are transformed to each other by time-reversal operator. In terms of the electron operators, these bound modes take the form
\begin{eqnarray}\label{eqn:}
\gamma_1&=&\sum_j\bigr[u^{(1)}_{\uparrow}(x_j)c_{\uparrow}(x_j)+u^{(1)}_{\downarrow}(x_j)c_{\downarrow}(x_j)+u^{(1)*}_{\uparrow}(x_j)c^\dag_{\uparrow}(x_j) +u^{(1)*}_{\downarrow}(x_j)c^\dag_{\downarrow}(x_j)\bigr],\\
\tilde\gamma_1&=&\sum_j\bigr[u^{(1)*}_{\uparrow}(x_j)c_{\downarrow}(x_j)-u^{(1)*}_{\downarrow}(x_j)c_{\uparrow}(x_j)+u^{(1)}_{\uparrow}(x_j)c^\dag_{\downarrow}(x_j) -u^{(1)}_{\downarrow}(x_j)c^\dag_{\uparrow}(x_j)\bigr],\\
\gamma_2&=&i\sum_j\bigr[u^{(2)}_{\uparrow}(x_j)c_{\uparrow}(x_j)+u^{(2)}_{\downarrow}(x_j)c_{\downarrow}(x_j)-u^{(2)*}_{\uparrow}(x_j)c^\dag_{\uparrow}(x_j) -u^{(2)*}_{\downarrow}(x_j)c^\dag_{\downarrow}(x_j)\bigr],\\
\tilde\gamma_2&=&i\sum_j\bigr[u^{(2)*}_{\uparrow}(x_j)c_{\downarrow}(x_j)-u^{(2)*}_{\downarrow}(x_j)c_{\uparrow}(x_j)-u^{(2)}_{\uparrow}(x_j)c^\dag_{\downarrow}(x_j) +u^{(2)}_{\downarrow}(x_j)c^\dag_{\uparrow}(x_j)\bigr].
\end{eqnarray}
Note that the coefficients in $H_{\rm wire}^{\rm eff}$ are real, and we have that $u^{(1,2)}_{\uparrow,\downarrow}=u^{(1,2)*}_{\uparrow,\downarrow}$. The coupling energies between the Majorana modes at left ($\gamma_{1},\tilde\gamma_1$) and right ($\gamma_2,\tilde\gamma_2$) ends are calculated by $E_1=i\langle\gamma_1|H_{\rm wire}^{\rm eff}|\gamma_2\rangle=-i\langle\tilde\gamma_1|H_{\rm wire}^{\rm eff}|\tilde\gamma_2\rangle$ and $E_2=i\langle\gamma_1|H_{\rm wire}^{\rm eff}|\tilde\gamma_2\rangle=-i\langle\tilde\gamma_1|H_{\rm wire}^{\rm eff}|\gamma_2\rangle$. Using the relations
\begin{eqnarray}\label{eqn:}
c_{j\uparrow}&\simeq&u^{(1)}_{\uparrow}(x_j)\gamma_1-u^{(1)}_{\downarrow}(x_j)\tilde\gamma_1-iu^{(2)}_{\uparrow}(x_j)\gamma_2+iu^{(2)}_{\downarrow}(x_j)\tilde\gamma_2,\\
c_{j\downarrow}&\simeq&u^{(1)}_{\downarrow}(x_j)\gamma_1+u^{(1)}_{\uparrow}(x_j)\tilde\gamma_1-iu^{(2)}_{\downarrow}(x_j)\gamma_2 -iu^{(2)}_{\uparrow}(x_j)\tilde\gamma_2,
\end{eqnarray}
we obtain that
\begin{eqnarray}\label{eqn:}
E_1&=&\sum_{\langle i,j\rangle\sigma}t_{ij}u^{(1)}_{\sigma}(x_i)u^{(2)}_{\sigma}(x_j)+\sum_{\langle i,j\rangle\sigma}\Delta_{ij}^pu^{(1)}_{\sigma}(x_i)u^{(2)}_{\sigma}(x_j)+\sum_{\langle i,j\rangle}t_{ij}^{\rm so}\bigr[u^{(1)}_{\uparrow}(x_i)u^{(2)}_{\downarrow}(x_j)+u^{(1)}_{\downarrow}(x_j)u^{(2)}_{\uparrow}(x_i)\bigr]\nonumber\\
&&+\sum_{j}\Delta_s\bigr[u^{(1)}_{\uparrow}(x_j)u^{(2)}_{\downarrow}(x_j)-u^{(1)}_{\downarrow}(x_j)u^{(2)}_{\uparrow}(x_j)\bigr]+\sum_{j,\sigma}V_j^{\rm dis}u^{(1)}_{\sigma}(x_j)u^{(2)}_{\sigma}(x_j),\\
E_2&=&\sum_{\langle i,j\rangle}t_{ij}\bigr[u^{(1)}_{\uparrow}(x_i)u^{(2)}_{\downarrow}(x_j)-u^{(1)}_{\downarrow}(x_i)u^{(2)}_{\uparrow}(x_j)\bigr]
+\sum_{\langle i,j\rangle}\Delta_{ij}^p\bigr[u^{(1)}_{\uparrow}(x_i)u^{(2)}_{\downarrow}(x_j)-u^{(1)}_{\downarrow}(x_i)u^{(2)}_{\uparrow}(x_j)\bigr]\nonumber\\
&&+\sum_{\langle i,j\rangle}t_{ij}^{\rm so}\bigr[u^{(1)}_{\uparrow}(x_i)u^{(2)}_{\uparrow}(x_j)-u^{(1)}_{\downarrow}(x_j)u^{(2)}_{\downarrow}(x_i)\bigr] +\sum_{j}\Delta_s\bigr[u^{(1)}_{\uparrow}(x_j)u^{(2)}_{\uparrow}(x_j)+u^{(1)}_{\downarrow}(x_j)u^{(2)}_{\downarrow}(x_j)\bigr]\nonumber\\
&&+\sum_{j,\sigma}V_j^{\rm dis}\bigr[u^{(1)}_{\uparrow}(x_j)u^{(2)}_{\downarrow}(x_j)-u^{(1)}_{\downarrow}(x_j)u^{(2)}_{\uparrow}(x_j).
\end{eqnarray}
Note that $t_{ij}^{\rm so}=-t_{ji}^{\rm so}$ due to time-reversal symmetry and for a uniform nanowire we have that $\sum_{\langle i,j\rangle}u^{(1)}_{\uparrow}(x_i)u^{(2)}_{\downarrow}(x_j)=\sum_{\langle i,j\rangle}u^{(2)}_{\uparrow}(x_i)u^{(1)}_{\downarrow}(x_j)$ and $\sum_{j}u^{(1)}_{\uparrow}(x_j)u^{(2)}_{\downarrow}(x_j)=\sum_{j}u^{(2)}_{\uparrow}(x_j)u^{(1)}_{\downarrow}(x_j)$. With these properties we find that in $E_1$ the terms corresponding to $t_{ij}^{\rm so}$ and $\Delta_s$ vanish, while in $E_2$ the terms for $t_{ij}, \Delta_p$, and $V_j^{\rm dis}$ vanish. We then have
\begin{eqnarray}\label{eqn:}
E_1&=&\sum_{\langle i,j\rangle\sigma}t_{ij}u^{(1)}_{\sigma}(x_i)u^{(2)}_{\sigma}(x_j)+\sum_{\langle i,j\rangle\sigma}\Delta_{ij}^pu^{(1)}_{\sigma}(x_i)u^{(2)}_{\sigma}(x_j)+\sum_{j,\sigma}V_j^{\rm dis}u^{(1)}_{\sigma}(x_j)u^{(2)}_{\sigma}(x_j),\\
E_2&=&\sum_{\langle i,j\rangle\sigma}t_{ij}^{\rm so}u^{(1)}_{\sigma}(x_i)u^{(2)}_{\sigma}(x_j) +\sum_{j,\sigma}\Delta_su^{(1)}_{\sigma}(x_j)u^{(2)}_{\sigma}(x_j).
\end{eqnarray}

The wave functions of Majorana bound modes decay exponentially as a function of the distance from the end of the nanowire, multiplying by an oscillatory function with the oscillating period equal to the Fermi wavelength in the nanowire. This implies that $u_{\sigma}^{(1)}\propto\sin(k_Fx)e^{-x/\xi}$ and $u_{\sigma}^{(2)}\propto\sin[k_F(L-x)]e^{-(L-x)/\xi}$, where $\xi$ is the effective coherence length of the wire. In the realistic material, we consider that the chemical potential in the nanowire is far below the half-filling condition and thus $k_Fa\ll1$. In this way we have $u^{(1)}_{\sigma}(x_j)u^{(2)}_{\sigma}(x_j)\approx u^{(1)}_{\sigma}(x_j)u^{(2)}_{\sigma}(x_{j\pm1})e^{\mp a/\xi}$. Furthermore, the coherence length (in the order of $1.0\mu$m) is typically much larger than the lattice constant $\xi\gg a$ ($a\sim0.5$nm), and we can further approximate that $u^{(1)}_{\sigma}(x_j)u^{(2)}_{\sigma}(x_j)\approx u^{(1)}_{\sigma}(x_j)u^{(2)}_{\sigma}(x_{j\pm1})$. Bearing this result in mind we get
\begin{eqnarray}\label{eqn:}
E_1&=&\sum_{\langle i,j\rangle\sigma}t_{ij}u^{(1)}_{\sigma}(x_i)u^{(2)}_{\sigma}(x_j)+\sum_{\langle i,j\rangle\sigma}\Delta_{ij}^pu^{(1)}_{\sigma}(x_i)u^{(2)}_{\sigma}(x_j)+\sum_{j,\sigma}V_j^{\rm dis}u^{(1)}_{\sigma}(x_j)u^{(2)}_{\sigma}(x_j),\\
E_2&=&\sum_{\langle i,j\rangle\sigma}t_{ij}^{\rm so}u^{(1)}_{\sigma}(x_i)u^{(2)}_{\sigma}(x_j)+\sum_{\langle i,j\rangle\sigma}\Delta_su^{(1)}_{\sigma}(x_i)u^{(2)}_{\sigma}(x_j).
\end{eqnarray}
The spin-orbit hopping coefficient $t_{ij}^{\rm so}=-t_{ji}^{\rm so}$ and the $p$-wave pairing $\Delta_{ij}^p=-\Delta_{ji}^p$ are staggered parameters. In the limit that $k_Fa\ll1$ and $\xi\gg a$, the summation for such two terms in $E_1$ and $E_2$ also turns out to be zero. On the other hand, the spin-conserved hopping is a constant and we denote $t_{ij}=t_{ji}=t$. Finally, if the the random potential $V_j^{\rm dis}$ with $\langle V_j^{\rm dis}\rangle=0$ is distributed homogeneously in the nanowire, we expect that the last term in $E_1$ gives $V_0\sum_{j,\sigma}u^{(1)}_{\sigma}(x_j)u^{(2)}_{\sigma}(x_j)$ with the constant factor $V_0$ depending on the specific disorder profile and much less than the amplitude of the disorder potential. The couplings $E_{1,2}$ become
\begin{eqnarray}
E_1&\simeq&(t+V_0)\sum_{\langle i,j\rangle\sigma}u^{(1)}_{\sigma}(x_i)u^{(2)}_{\sigma}(x_j),\label{eqn:coupling1}\\
E_2&\simeq&\Delta_s\sum_{\langle i,j\rangle\sigma}u^{(1)}_{\sigma}(x_i)u^{(2)}_{\sigma}(x_j).\label{eqn:coupling2}
\end{eqnarray}
From the above result we find that $E_2/E_1\approx\Delta_s/(t+V_0)$, which is consistent with the fact that when $\Delta_s=0$ the original Hamiltonian~\eqref{eqn:nanowiretightbinding} can be block diagonalized and then $E_2\equiv0$. This implies that in the realistic nanowire materials while the magnitudes of $E_{1,2}(\lambda)$ depend on $\lambda$ which determines the overlapping between the wave functions of Majorana bound modes at left and right ends, their ratio is nearly a constant. Therefore we always have
\begin{eqnarray}\label{eqn:condition3}
\partial_\lambda\theta\approx0,
\end{eqnarray}
\begin{figure}[h]
\includegraphics[width=0.75\columnwidth]{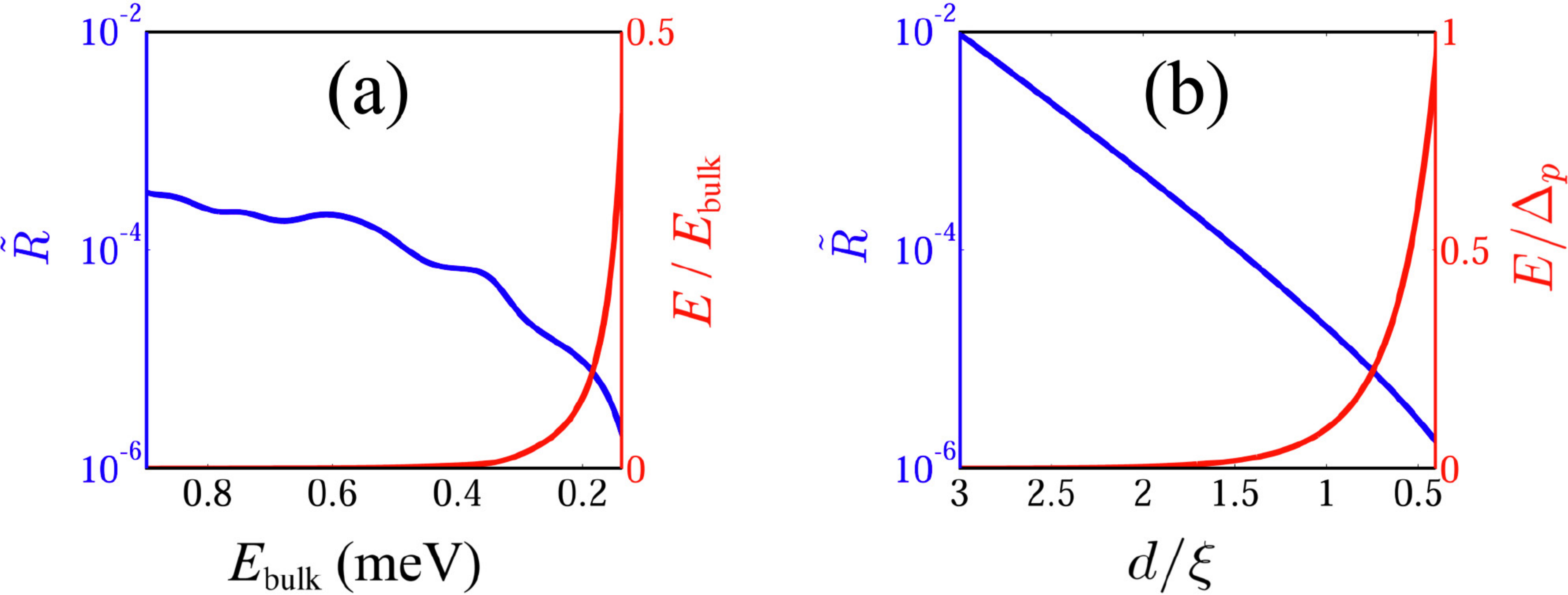}\caption{\small Adiabatic condition and fermion parity conservation for each sector of time-reversal partners in the presence of disorder scattering. The energy splitting $E$ between $|0\tilde0\rangle$ and $|1\tilde1\rangle$ (red curves) and the ratio $\tilde R$ (blue curves) versus (a) the bulk gap which varies by tuning the chemical potential, and (b) the distance between the Majorana end modes. In the numerical simulation the random on-site disorder potential is considered, with the potential amplitude $V_{\rm dis}\sim1.0$meV. Other parameters in the nanowire are taken that $\Delta_p=1.0$meV, $\Delta_s=0.5$meV, and $E_{\rm so}=0.1$meV. The coupling energy $E$ is tuned from $0$ to $1.0$meV within the time $1.0\mu$s.}
\label{SIadiabatic}
\end{figure}
which validates the adiabatic condition. The results in the Eqs.~\eqref{eqn:coupling1} and \eqref{eqn:coupling2} can be interpreted by an intuitive physical picture. Being proportional to the overlapping between the wave functions of Majorana bound modes at different ends, the coupling coefficients $E_{1,2}$ are exponential decaying functions of the nanowire length. Since the Majorana modes $\gamma_j$ and $\tilde\gamma_j$ are connected by ${\cal T}$-transformation, their wave functions have exactly the same spatial profile, which leads to the same exponential form for the coefficients $E_{1,2}(\lambda)=\alpha_{1,2}(\lambda)e^{-d/\xi}$ with $d$ the distance between the left and right Majorana end modes. The pre-factors $\alpha_j(\lambda)$ depend on the local couplings, i.e. the hopping coefficients and pairings between electrons belonging to the same (for $j=1$) or different (for $j=2$) sectors of the time-reversal partners. For the case with constant and homogeneous local couplings, we have that their ratio $\alpha_1/\alpha_2$ is proportional to the ratio of couplings between electrons of the same and different sectors, and is nearly a constant, justifying the adiabatic condition. The above derivative is clearly confirmed with numerical results in the realistic systems with the presence of random on-site disorder scattering, as shown in Fig.~\ref{SIadiabatic}.

It is worthwhile to note that for fixed parameter $\lambda$, the physics of the fermion parity conservation for each sector can be easily understood in another way. For DIII class topological superconductor, the helical $p$-wave pairings occur between two electrons belonging to the same sector of the time-reversal partners. While the change by one in the fermion number of each sector conserves the total fermion parity of the system, it changes fermion parity for each sector, and thus breaks a $p$-wave Cooper pair in each sector. This process costs finite energy and is thus suppressed by the $p$-wave pairing gap if $\Delta_p$ dominates over $\Delta_s$ and the time-reversal symmetry is not broken. The previous study in this section further proves this conservation law when the couplings between Majorana end modes are allowed and adjusted adiabatically. To simplify the notations in the further discussion, we relabel the block diagonal Majorana modes $\gamma_j^{(1)},\tilde\gamma_j^{(1)}$ as $\gamma_j,\tilde\gamma_j$. Accordingly, the diagonal complex fermion modes are redefined as $f_j,\tilde f_j$.

\section{Josephson effect in the DIII class 1D topological superconductor}

Now we study how to measure the topological qubit states with the Josephson effect. It has been predicted that in the chiral 1D topological superconductor the Josephson current has $4\pi$ periodicity \cite{KitaevApp}, and the topological qubit states for a single wire, $|0\rangle$ and $|1\rangle$, can be read out from the direction of Josephson currents in the junction \cite{AliceaApp,FuApp}. In this section we predict a novel phenomenon in the Josephson effect of the DIII class 1D topological superconductor, which provides a feasible scheme to read out the topological qubit states in a TRI Majorana quantum wire.

\subsection{Effective coupling Hamiltonian}

We consider a Josephson junction formed by two Majorana nanowire ends with a phase difference $\phi=\phi_R-\phi_L$, as illustrated in Fig.~\ref{SIcurrent}(a), and derive the effective coupling Hamiltonian for the Majorana zero modes localized at the left $(L)$ and right $(R)$ ends. The electron tunneling process in the junction is described by
\begin{eqnarray}\label{eqn:junction1}
H_T=\Upsilon c_{L,N}^\dag c_{R,1}+\Upsilon\tilde c_{L,N}^\dag\tilde c_{R,1}+{\rm H.c.},
\end{eqnarray}
where $c_{L,N}, \tilde c_{L,N}$ and $c_{R,1}, \tilde c_{R,1}$ represent the electron operators for the $N$th site at left and $1$st site at right ends of the junction, respectively, and $\Upsilon$ is the tunneling coefficient across the junction. The Majorana end modes can be generically expanded in terms of electron operators
\begin{eqnarray}
\gamma_L&=&\sum_{j}\bigr(u_{L,j}c_{L,j}+u_{L,j}^*c_{L,j}^\dag\bigr), \ \tilde\gamma_L=\sum_{j}\bigr(\tilde u_{L,j}\tilde c_{L,j}+\tilde u_{L,j}^*\tilde c_{L,j}^\dag\bigr),\\
\gamma_R&=&\sum_{j}\bigr(u_{R,j}c_{R,j}+u_{R,j}^*c_{R,j}^\dag\bigr), \ \tilde\gamma_R=\sum_{j}\bigr(\tilde u_{R,j}\tilde c_{R,j}+\tilde u_{R,j}^*\tilde c_{R,j}^\dag\bigr),
\end{eqnarray}
where $u_{L/R,j}=\tilde u^*_{L/R,j}$ if $\phi_{L/R}=0$.
Note that $c_{j}$ and $\tilde c_j$ represent electron operators of a general time-reversal pair at $j$th site, not necessarily corresponding to spin-up and spin-down, since the spin is not a good quantum number when spin-orbit coupling and $s$-wave order are present. From the above formulas we can solve the electron operators in terms of Majorana and nonzero energy Bogoliubov quasiparticle operators. Reexpressing the Bogoliubov quasiparticles in terms of electron operators we can interpret $c_{L,N},\tilde c_{L,N}$ and $c_{R,1},\tilde c_{R,1}$ by
\begin{eqnarray}
c_{L,N}&=&u^*_{L,N}\gamma_L-\sum_{j=1}^{N}a_{L,j}c_{L,j}-\sum_{j=1}^Nb_{L,j}^*c_{L,j}^\dag,\label{eqn:junction2} \\
\tilde c_{L,N}&=&\tilde u^*_{L,N}\gamma_L-\sum_{j=1}^{N}\tilde a_{L,j}\tilde c_{L,j}-\sum_{j=1}^N\tilde b_{L,j}^*\tilde c_{L,j}^\dag, \\
c_{R,1}&=&u^*_{R,1}\gamma_R-\sum_{j=1}^Na_{R,j}c_{R,j}-\sum_{j=1}^Nb_{R,j}^*c_{R,j}^\dag,\\
\tilde c_{R,1}&=&\tilde u^*_{R,1}\gamma_R-\sum_{j=1}^N\tilde a_{R,j}\tilde c_{R,j}-\sum_{j=1}^N\tilde b_{R,j}^*\tilde c_{R,j}^\dag,\label{eqn:junction2'}
\end{eqnarray}
with a constant normalization factor neglected. Here $a_{L/R,j},\tilde a_{L/R,j}$ and $b_{L/R,j},\tilde b_{L/R,j}$ are expansion coefficients, originated from the quasiparticle operators other than the corresponding Majorana mode.
Substituting these results into the tunneling Hamiltonian $H_T$ yields that
\begin{eqnarray}\label{eqn:junction3}
H_T&=&\Upsilon\biggr(u_{L,N}\gamma_L-\sum_{j=1}^{N}a^*_{L,j}c^\dag_{L,j}-\sum_{j=1}^Nb_{L,j}c_{L,j}\biggr) \biggr(u^*_{R,1}\gamma_R-\sum_{j=1}^Na_{R,j}c_{R,j}-\sum_{j=1}^Nb_{R,j}^*c_{R,j}^\dag\biggr)+\nonumber\\
&&+\Upsilon\biggr(\tilde u_{L,N}\tilde\gamma_L-\sum_{j=1}^{N}\tilde a^*_{L,j}\tilde c^\dag_{L,j}-\sum_{j=1}^N\tilde b_{L,j}\tilde c_{L,j}\biggr)\biggr(\tilde u^*_{R,1}\tilde\gamma_R-\sum_{j=1}^N\tilde a_{R,j}\tilde c_{R,j}-\sum_{j=1}^N\tilde b_{R,j}^*\tilde c_{R,j}^\dag\biggr)+{\rm H.c.}\nonumber\\
&\approx&H^{(0)}+H^{(1)},
\end{eqnarray}
where
\begin{eqnarray}
H^{(0)}&=&\Upsilon u_{L,N}u^*_{R,1}\gamma_L\gamma_R+\Upsilon \tilde u_{L,N}\tilde u^*_{R,1}\tilde\gamma_L\tilde\gamma_R+{\rm H.c.},\nonumber\\
H^{(1)}&=&-\Upsilon u_{L,N}\gamma_L\biggr(\sum_{j=1}^Na_{R,j}c_{R,j}+\sum_{j=1}^Nb_{R,j}^*c_{R,j}^\dag\biggr)
-\Upsilon u_{R,1}\gamma_R\biggr(\sum_{j=1}^{N}a_{L,j}c_{L,j}+\sum_{j=1}^Nb^*_{L,j}c^\dag_{L,j}\biggr)\nonumber\\
&&-\Upsilon \tilde u_{L,N}\tilde\gamma_L\biggr(\sum_{j=1}^N\tilde a_{R,j}\tilde c_{R,j}+\sum_{j=1}^N\tilde b_{R,j}^*\tilde c_{R,j}^\dag\biggr)
-\Upsilon \tilde u_{R,1}\tilde\gamma_R\biggr(\sum_{j=1}^{N}\tilde a_{L,j}\tilde c_{L,j}+\sum_{j=1}^N\tilde b^*_{L,j}\tilde c^\dag_{L,j}\biggr)+{\rm H.c.}.\nonumber
\end{eqnarray}
In the second equation of the formula~\eqref{eqn:junction3} we have neglected the higher-order irrelevant terms. The term $H^{(0)}$ represents the direct coupling between Majorana modes at different junction ends, which gives the first term of the effective Hamiltonian $H_{\rm eff}$ in the main text. This can be seen by noticing that
\begin{equation}\label{eqn:wavefunction}
\begin{split}
u_{L,N}&=i|u_{L,N}|e^{i\phi_L/2}, \ u_{R,1}=|u_{R,1}|e^{i\phi_R/2},\\
\tilde u_{L,N}&=-i|\tilde u_{L,N}|e^{i\phi_L/2}, \ \tilde u_{R,1}=|\tilde u_{R,1}|e^{i\phi_R/2},
\end{split}
\end{equation}
with which we can recast $H^{(0)}$ into
\begin{eqnarray}
H^{(0)}=i\Gamma_0\cos\frac{\phi}{2}\bigr(\gamma_L\gamma_R-\tilde\gamma_L\tilde\gamma_R\bigr), \ \Gamma_0=2\Upsilon|u_{L,N}u_{R,1}|.
\end{eqnarray}

On the other hand, for $H^{(1)}$ we shall calculate up to the second-order perturbation, which is responsible for the second term of $H_{\rm eff}$ in the main text. From $H^{(1)}$ we know that Majorana modes at one end (e.g. the left end) also couple to the electron modes at another end (the right end). In the second-order perturbation $\gamma_L$ and $\tilde\gamma_L$ ($\gamma_R$ and $\tilde\gamma_R$) couple to $c_R$ and $\tilde c_R$ ($c_L$ and $\tilde c_L$), respectively. When a nonzero $s$-wave pairing is present in the quantum wires, the electrons $c_{L/R}$ and $\tilde c_{L/R}$ form a Cooper pair and condense. This process leads to an effective coupling between Majorana zero modes localized at the same end. Therefore, up to the second-order perturbation in the tunneling process, we obtain that
\begin{eqnarray}
H^{(1)}_{\rm eff}&=&\frac{1}{2}\Upsilon^2u_{L,N}\tilde u_{L,N}\gamma_L\tilde\gamma_L\biggr[\sum_{j=1}^Na_{R,j}\tilde a_{R,j}\int d\tau\langle T_\tau c_{R,j}(\tau)\tilde c_{R,j}(0)\rangle+\sum_{j=1}^Nb_{R,j}^*\tilde b_{R,j}^*\int d\tau\langle T_\tau c_{R,j}^\dag(\tau)\tilde c^\dag_{R,j}(0)\rangle\biggr]\nonumber\\
&&+\frac{1}{2}\Upsilon^2u_{R,1}\tilde u_{R,1}\gamma_R\tilde\gamma_R\biggr[\sum_{j=1}^{N}a_{L,j}\tilde a_{L,j}\int d\tau\langle T_\tau c_{L,j}(\tau)\tilde c_{L,j}(0)\rangle+\sum_{j=1}^Nb_{L,j}^*\tilde b_{L,j}^*\int d\tau\langle c_{L,j}^\dag(\tau)\tilde c^\dag_{L,j}(0)\rangle\biggr]+{\rm H.c.}
\end{eqnarray}
Here $\int d\tau\langle T_\tau \cdots\rangle$ represents time ordered integral. Assuming that the superconducting pairings are uniform in the Majorana nanowires, we obtain from the above formula that
\begin{eqnarray}
H^{(1)}_{\rm eff}&=&\frac{1}{2}\Upsilon^2u_{L,N}\tilde u_{L,N}\gamma_L\tilde\gamma_L\biggr[\sum_{j=1}^Na_{R,j}\tilde a_{R,j}\sum_k\frac{\Delta_{s,R}^*}{{\cal E}_R^2(\Delta_{s,R};\Delta_{p,R};k)}-\sum_{j=1}^Nb_{R,j}^*\tilde b_{R,j}^*\sum_k\frac{\Delta_{s,R}}{{\cal E}_R^2(\Delta_{s,R};\Delta_{p,R};k)}\biggr]\nonumber\\
&&+\frac{1}{2}\Upsilon^2u_{R,1}\tilde u_{R,1}\gamma_R\tilde\gamma_R\biggr[\sum_{j=1}^{N}a_{L,j}\tilde a_{L,j}\sum_k\frac{\Delta_{s,L}^*}{{\cal E}_L^2(\Delta_{s,L};\Delta_{p,L};k)}-\sum_{j=1}^Nb_{L,j}^*\tilde b_{L,j}^*\sum_k\frac{\Delta_{s,L}}{{\cal E}_L^2(\Delta_{s,L};\Delta_{p,L};k)}\biggr]+{\rm H.c.}\nonumber\\
&=&i\Upsilon_L(\phi)\gamma_L\tilde\gamma_L+i\tilde\Upsilon_R(\phi)\gamma_R\tilde\gamma_R,
\end{eqnarray}
with ${\cal E}_{L/R}(\Delta_{s,R};\Delta_{p,R};k)$ the bulk excitation spectra in the left (for $L$) and right (for $R$) wires of the junction, respectively. The coupling coefficients read
\begin{eqnarray}
\Upsilon_{L/R}(\phi)&=&-i\frac{1}{2}\Upsilon^2u_{L/R,N/1}\tilde u_{L/R,N/1}\biggr[\sum_{j=1}^Na_{R/L,j}\tilde a_{R/L,j}\sum_k\frac{\Delta_{s,R/L}^*}{{\cal E}_{R/L}^2(\Delta_{s,R/L};\Delta_{p,R/L};k)}\nonumber\\
&&-\sum_{j=1}^Nb_{R/L,j}^*\tilde b_{R/L,j}^*\sum_k\frac{\Delta_{s,R/L}}{{\cal E}_{R/L}^2(\Delta_{s,R/L};\Delta_{p,R/L};k)}\biggr]-{\rm c.c.}
\end{eqnarray}
With the relations obtained in Eqs.~\eqref{eqn:junction2} to \eqref{eqn:junction2'} we have that $a_{R/L,j}\tilde a_{R/L,j}=|a_{R/L,j}\tilde a_{R/L,j}|$, and
$b_{R/L,j}\tilde b_{R/L,j}=|b_{R/L,j}\tilde b_{R/L,j}|e^{i2\phi_{R/L}}$.
Together with the results in Eq.~\eqref{eqn:wavefunction} we can simplify $\Upsilon_{L/R}(\phi)$ to be
\begin{eqnarray}
\Upsilon_{L}(\phi)&=&-i\Gamma_1e^{i\phi}-{\rm c.c.}=\Gamma_1\sin\phi,\nonumber\\
\Upsilon_{R}(\phi)&=&-i\Gamma_1e^{-i\phi}-{\rm c.c.}=-\Gamma_1\sin\phi,\nonumber
\end{eqnarray}
and the effective coupling Hamiltonian for Majorana fermions at the same end takes the following form
\begin{eqnarray}
H^{(1)}_{\rm eff}=i\Gamma_1\sin\phi(\gamma_L\tilde\gamma_L-\gamma_R\tilde\gamma_R).
\end{eqnarray}
The coupling constant $\Gamma_1$ is calculated by
\begin{eqnarray}
\Gamma_1=\Upsilon^2|u_{L,N}\tilde u_{L,N}|\biggr[\sum_{j=1}^N|a_{R,j}\tilde a_{R,j}|\sum_k\frac{|\Delta_{s,R}|}{{\cal E}_{R}^2(\Delta_{s,R};\Delta_{p,R};k)}-\sum_{j=1}^N|b_{R,j}^*\tilde b_{R,j}^*|\sum_k\frac{|\Delta_{s,R}|}{{\cal E}_{R}^2(\Delta_{s,R};\Delta_{p,R};k)}\biggr].
\end{eqnarray}
We have assumed the uniformity of the parameters in the left and right wires of the junction that $|\Delta_{s,L}|=|\Delta_{s,R}|, |\Delta_{p,L}|=|\Delta_{p,R}|$, and therefore $|u_{L,N}\tilde u_{L,N}|=|u_{R,1}\tilde u_{R,1}|$ and ${\cal E}_{L}(\Delta_{s,L};\Delta_{p,L};k)={\cal E}_{R}(\Delta_{s,R};\Delta_{p,R};k)$. We note that this condition is typically satisfied in the realistic systems. It is clear that the $\Gamma_1$-term vanishes when the $s$-wave pairing $\Delta_{s,L/R}$ is absent in the wires.

To this end, we combine $H^{(0)}$ and $H^{(1)}_{\rm eff}$ to reach finally the effective Hamiltonian for a Josephson junction formed by DIII class topological superconductors that
\begin{eqnarray}\label{eqn:effective1}
H_{\rm eff}(\phi)=i\Gamma_0\cos\frac{\phi}{2}(\gamma_L\gamma_R-\tilde\gamma_L\tilde\gamma_R)+i\Gamma_1\sin\phi(\gamma_L\tilde\gamma_L-\gamma_R\tilde\gamma_R).
\end{eqnarray}
Note that if treating $\phi$ as a fixed parameter, the $\Gamma_1$-term in the above formula breaks time-reversal symmetry. This reflects that the leading-order
contribution to the coupling between Majoranas at the same end ($i\gamma_j\tilde\gamma_j$) should come from the second-order perturbation in the tunneling process. Actually, the direct coupling between $\gamma_j$ and $\tilde\gamma_j$ does not experience the phase difference across the junction and should preserve time-reversal symmetry. This is because a uniform pairing phase in one end of the junction can be removed by a constant gauge transformation. Therefore the coupling between Majorana fermions at the same end can only be induced by electron tunneling across the junction and the minimum requirement is to consider the second-order tunneling process. Furthermore, the system restores time-reversal symmetry at $\phi=m\pi$, which explains why the $\Gamma_1$-term is proportional to $\sin\phi$, and has $2\pi$ periodicity.

\subsection{Josephson current}

\begin{figure}[ht]
\includegraphics[width=0.6\columnwidth]{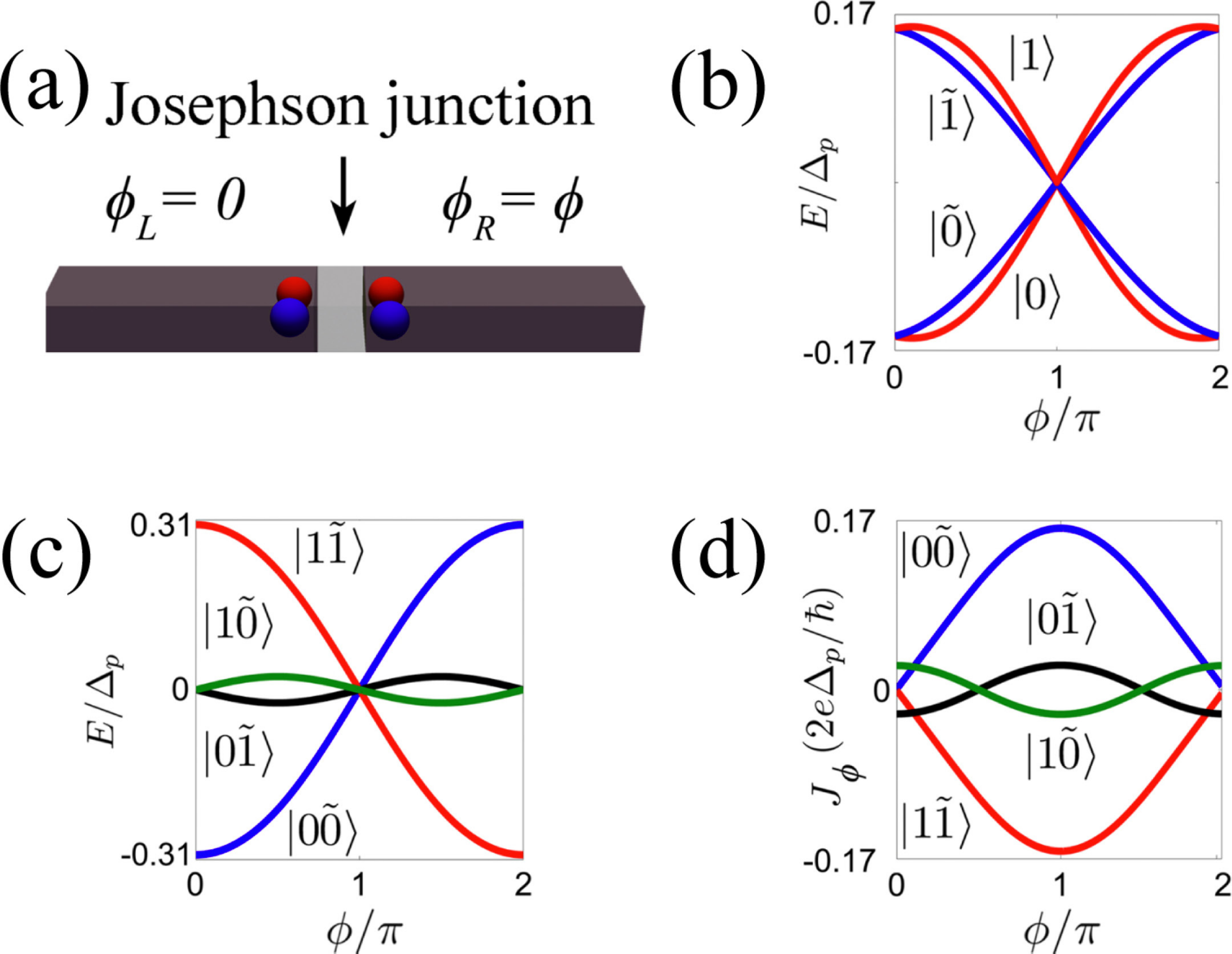}\caption{\small Josephson effect in DIII class 1D topological superconductor with the inclusion of random disorder scattering. (a) The sketch of a Josephson junction with phase difference $\phi$. (b) The single particle Andreev bound state spectra versus the phase difference $\phi$. (c) The energy spectra of the four qubit states $|n_1\tilde n_1\rangle$ ($n_1,\tilde n_1=0,1$) according to the results in (b). (d) The Josephson currents (in units of $2e\Delta_p/\hbar$) for different topological qubit states. In the numerical simulation the amplitude of the random on-site disorder potential is set as $V_{\rm dis}\sim1.0$meV. Other parameters are taken that $\Delta_p=1.0$meV, $\Delta_s=0.25$meV, $E_{\rm so}=0.1$meV, the width of the junction $d=0.75\xi$, and in middle trivial region (gray color) of the junction the chemical potential is set to be at the band bottom.}
\label{SIcurrent}
\end{figure}

The Hamiltonian~\eqref{eqn:effective1} can be block diagonalized by a constant transformation in the Majorana bases that $\gamma'_1=\gamma_L+\tilde\gamma_R$, $\gamma'_2=\gamma_R+\tilde\gamma_L$, and $\tilde\gamma'_{1,2}={\cal T}\gamma'_{1,2}{\cal T}^{-1}$, which sends $H_{\rm eff}$ to be $H_{\rm eff}=i(\Gamma_0\cos\phi/2+\Gamma_1\sin\phi)\gamma'_1\gamma'_2-i(\Gamma_0\cos\phi/2-\Gamma_1\sin\phi)\tilde\gamma'_1\tilde\gamma'_2$. The Andreev bound state spectra are obtained straightforwardly by $E_{n_{f'},\tilde n_{f'}}(\phi)=\bigr(\Gamma_0\cos\phi/2+\Gamma_1\sin\phi\bigr)(2n_{f'}-1)+\bigr(\Gamma_0\cos\phi/2-\Gamma_1\sin\phi\bigr)(2n_{\tilde f'}-1)$,
which are doubly degenerate at $\phi=m\pi$, reflecting the time-reversal symmetry at these points. Here $n_{f',\tilde f'}$ are complex fermion number operators for $f'$ and $\tilde f'$ modes, respectively. The Josephson current then reads
\begin{eqnarray}\label{eqn:Josephson2}
J_{\phi}=\bigr(\frac{e\Gamma_0}{2\hbar}\sin\frac{\phi}{2}+\frac{e\Gamma_1}{2\hbar}\cos\phi\bigr)(2n_{f'}-1)+\bigr(\frac{e\Gamma_0}{2\hbar}\sin\frac{\phi}{2}-\frac{e\Gamma_1}{2\hbar}\cos\phi\bigr)(2n_{\tilde f'}-1).
\end{eqnarray}

From the equation~\eqref{eqn:Josephson2} we find that for the even parity states ($|0\tilde0\rangle$ and $|1\tilde1\rangle$) the Josephson currents $J^{\rm even}_\phi=\pm\frac{e}{\hbar}\Gamma_0\sin\frac{\phi}{2}$, which are of $4\pi$ periodicity, while for the odd parity states ($|0\tilde1\rangle$ and $|1\tilde0\rangle$) $J^{\rm odd}_\phi=\pm\frac{e}{\hbar}\Gamma_1\cos\phi$ exhibit $2\pi$ periodicity. The difference in the periodicity reflects different mechanisms for $J^{\rm even,\rm odd}_\phi$. The currents $J^{\rm even}_\phi$ are contributed from the $\Gamma_0$-term in the effective coupling Hamiltonian, which is due to the direct coupling between Majorana modes at different ends of the junction. Therefore the currents $J^{\rm even}_\phi$ are a consequence of the single-electron tunneling process and has $4\pi$ periodicity. On the other hand, as contributed from the $\Gamma_1$ term, the Josephson currents $J^{\rm odd}_\phi$ are resulted from the second-order tunneling process which corresponds to the Cooper pair tunneling, therefore being of $2\pi$ periodicity. Furthermore, the currents $J^{\rm odd}_\phi$ are nonzero even for $\phi=0$, which reflects the fact that the odd-parity states violate time-reversal symmetry even $H_{\rm eff}$ preserves at $\phi=m\pi$.

The $4\pi$ periodicity of the Josephson currents for even parity states can also be understood in the following way. When the phase difference across the junction advances $2\pi$, the Cooper pair wave function changes $2\pi$ across the junction, while for single electron operators the phase varies only $\pi$. This implies that the coupling coefficients also change $\pi$ phase and thus reverse sign, leading to the $4\pi$ periodicity of the direct coupling term.
The generality of this argument implies that the $4\pi$ periodicity of $J^{\rm even}_\phi$ is stable against the disorder scattering without breaking time-reversal symmetry. On the other hand, for odd parity states, the two-fold degeneracy at $\phi=m\pi$ is protected by time-reversal symmetry, which shows that the qualitative properties of the Josephson currents $J^{\rm odd}_\phi$ are also stable against the TRI disorder scattering. The numerical results are shown in Fig.~\ref{SIcurrent}.

With the above results we can have different strategies in the experiment to distinguish $J^{\rm even}_\phi$ and $J^{\rm odd}_\phi$. For instance, one can measure the periodicity of the Josephson currents, or measure the currents at $\phi=\pi/2$ where $J^{\rm even}_\phi=\pm\frac{e}{\sqrt{2}\hbar}\Gamma_0$ and $J^{\rm odd}_\phi=0$. Furthermore, the two qubit states of the same total parity are distinguished by current directions. The qualitative difference in the Josephson measurements provides direct detection of the four topological qubit states.


\noindent

\onecolumngrid

\renewcommand{\thesection}{S-\arabic{section}}
\renewcommand{\theequation}{S\arabic{equation}}
\setcounter{equation}{0}  
\renewcommand{\thefigure}{S\arabic{figure}}
\setcounter{figure}{0}  

\section*{\Large\bf Supplementary Material}

In this Supplementary Material we provide the details of the braiding statistics of Majorana end modes in DIII class 1D topological superconductor.

\section*{The general picture}

With the fermion parity conservation for each sector shown in the Appendix of the manuscript we can derive that the exchange of Majorana zero modes in DIII class 1D topological superconductor is generically equivalent to two independent processes of braiding Majorana fermions of two different sectors, respectively. The physics can be understood in the following way. First of all, since the bulk is gapped, braiding adiabatically the Majorana zero modes, e.g. $\gamma_1,\tilde\gamma_1$ and $\gamma_2,\tilde\gamma_2$, does not affect the bulk states. Furthermore, assuming that other Majorana zero modes are located far away from the two exchanged pairs, in the braiding we only need to consider the evolution of the Majorana zero modes which are braided. Finally, since the fermion parity is conserved for each sector, in the braiding the two complex fermion modes $f_1$ and $\tilde f_1$ are decoupled and their dynamics can be derived independently. This leads to the braiding matrix given in the manuscript and studied in detail in the following.

\subsection{Degenerate ground states}

We first construct the generic degenerate ground states for the DIII class 1D topological superconductor. Consider that the 1D Majorana wire has $2M$ pairs of Majorana zero modes $\gamma_1, \tilde\gamma_1; \gamma_2, \tilde\gamma_2; ...,$ and $\gamma_{2M}, \tilde\gamma_{2M}$, with different pairs of Majorana zero modes well separated from each other. With these modes we can define the $M$ pairs of complex fermion modes by $f_j=(\gamma_{2j-1}+i\gamma_{2j})/2$ and $\tilde f_j=(\tilde\gamma_{2j-1}-i\tilde\gamma_{2j})/2$, with $j=1,...,M$. It follows that
\begin{eqnarray}\label{eqn:complex1}
{\cal T}^{-1}f_j{\cal T}=\tilde f_j, \ \ {\cal T}^{-1}\tilde f_j{\cal T}=-f_j.
\end{eqnarray}

For the bulk states, we denote the Bogoliubov-de Gennes quasiparticle operators with positive energies by $d_\eta$ and $\tilde d_\eta$, with $\tilde d_\eta={\cal T}^{-1} d_\eta{\cal T}$. Let $|\rm vac\rangle$ be the vacuum state with respect to the electron operators and we construct the wave function
\begin{eqnarray}
|\psi\rangle=\frac{1}{\sqrt{\cal N}}\prod_{\alpha}d_\alpha\prod_\beta\tilde d_\beta|\rm vac\rangle.
\end{eqnarray}
Here $\sqrt{\cal N}$ is the normalization factor. It is clear that $|\psi\rangle$ is an eigenstate of $f_j^\dag f_j$ and $\tilde f_j^\dag\tilde f_j$ with eigenvalue $n_j$ and $\tilde n_j$, respectively. Note that $n_j$ and $\tilde n_j$ can be either $1$ or $0$, and their magnitudes depend on the convention used in defining the Majorana wave functions \cite{IvanovApp,AliceaApp}. To be concrete, we shall use the convention that
\begin{eqnarray}
|\psi\rangle=|11...1\rangle|\tilde 1\tilde 1...\tilde 1\rangle.
\end{eqnarray}
The ground state with $n_j=\tilde n_j=0$ is then constructed by
\begin{eqnarray}
|00...0\rangle|\tilde 0\tilde 0...\tilde 0\rangle=f_1...f_M\tilde f_1...\tilde f_1|\psi\rangle,
\end{eqnarray}
and generically
\begin{eqnarray}\label{eqn:ground1}
|n_1...n_M\rangle|\tilde n_1...\tilde n_M\rangle=\prod_{j}^Mf_j^{1-n_j}\prod_{j}^M\tilde f_j^{1-\tilde n_j}|\psi\rangle.
\end{eqnarray}

\subsection{General results for the braiding process}

Now we study the general results of the braiding process. In particular, we shall show that braiding two pairs of Majorana zero modes $\gamma_j,\tilde\gamma_j$ and $\gamma_{j+1},\tilde\gamma_{j+1}$ can be reduced to two independent processes of exchanging $\gamma_{j},\gamma_{j+1}$ and $\tilde\gamma_{j},\tilde\gamma_{j+1}$, respectively. During the braiding process, the Hamiltonian $H(\lambda)$ generically depends on an adiabatic parameter $\lambda$, and at any fixed $\lambda$-value the time-reversal symmetry is preserved. There are two different contributions which may affect the exchange dynamics. One is the Berry phase effect in the degenerate ground subspace, and another is that after the exchange, the original ground state subspace may vary and evolve into new forms. This study is similar as that in the chiral topological superconductors \cite{IvanovApp,AliceaApp}, but with the new ingredients of time-reversal symmetry and fermion parity conservation for each sector of the time-reversal partners.

\subsection*{B.1. Berry phase effect}

In the braiding, the Hamiltonian and the many-body ground state vary with the adiabatic parameter $\lambda$. To determine the Berry phase in the braiding, we calculate the Berry's connection by
\begin{eqnarray}
A_{n,m}(\lambda)=i\hbar\langle n_1...n_M|\langle\tilde n_1...\tilde n_M|\partial_\lambda|m_1...m_M\rangle|\tilde m_1...\tilde m_M\rangle.
\end{eqnarray}
Since we only braid the two pairs of Majorana zero modes $\gamma_{1},\tilde\gamma_{1}$ and $\gamma_{2},\tilde\gamma_{2}$, we can assume that only these four Majorana zero modes are $\lambda$-dependent, while all other Majorana modes are independent of $\lambda$ \cite{AliceaApp}. Then, from Eq.~\eqref{eqn:ground1} we know that
\begin{eqnarray}
A_{n,m}(\lambda)&=&A_{n_1\tilde n_1,m_1\tilde m_1}\delta_{n_2m_2}\delta_{\tilde n_2\tilde m_2}...\delta_{n_Mm_M}\delta_{\tilde n_M\tilde m_M},
\end{eqnarray}
where
\begin{eqnarray}
A_{n_1\tilde n_1,m_1\tilde m_1}(\lambda)&=&i\hbar\langle n_1;\tilde n_1|\partial_\lambda|m_1;\tilde m_1\rangle\nonumber\\
&=&A^{(1)}_{n_1\tilde n_1,m_1\tilde m_1}(\lambda)+A^{(2)}_{n_1\tilde n_1,m_1\tilde m_1}(\lambda)+A^{(3)}_{n_1\tilde n_1,m_1\tilde m_1}(\lambda).
\end{eqnarray}
For the last line of the above formula we have that
\begin{eqnarray}\label{eqn:Berry0}
A^{(1)}_{n_1\tilde n_1,m_1\tilde m_1}(\lambda)&=&i\hbar\langle\psi(\lambda)|[\tilde f_1^\dag]^{1-\tilde n_1}[f_1^\dag]^{1-n_1}[f_1]^{1-m_1}[\tilde f_1]^{1-\tilde m_1}\partial_\lambda |\psi(\lambda)\rangle,\nonumber\\
A^{(2)}_{n_1\tilde n_1,m_1\tilde m_1}(\lambda)&=&i\hbar(1-m_1)\langle\psi(\lambda)|[\tilde f_1^\dag]^{1-\tilde n_1}[f_1^\dag]^{1-n_1}(\partial_\lambda f_1)[\tilde f_1]^{1-\tilde m_1}|\psi(\lambda)\rangle,\\
A^{(3)}_{n_1\tilde n_1,m_1\tilde m_1}(\lambda)&=&i\hbar(1-\tilde m_1)\langle\psi(\lambda)|[\tilde f_1^\dag]^{1-\tilde n_1}[f_1^\dag]^{1-n_1}[f_1]^{1-m_1}(\partial_\lambda \tilde f_1)|\psi(\lambda)\rangle.\nonumber
\end{eqnarray}
From the former section we have shown that the Fermi parity is conserved for each sector of the time-reversal partners, and therefore we have $A_{n_1\tilde n_1,m_1\tilde m_1}=0$ for $n_1\neq m_1$ or $\tilde n_1\neq\tilde m_1$. On the other hand, the continuous variation of $\lambda$ cannot change the fermion numbers which are discrete values, which implies that
\begin{eqnarray}
\partial_\lambda |\psi(\lambda)\rangle\propto|\psi(\lambda)\rangle,
\end{eqnarray}
and therefore we have
\begin{eqnarray}
A^{(j)}_{n_1\tilde n_1,m_1\tilde m_1}(\lambda)=A_{n_1\tilde n_1}^{(j)}(\lambda)\delta_{n_1m_1}\delta_{\tilde n_1\tilde m_1}, \ j=1,2,3.
\end{eqnarray}
It is easy to check that $A_{n_1\tilde n_1}^{(1)}(\lambda)$ is independent of $n_1$ and $\tilde n_1$. To calculate $A^{(2)}_{n_1\tilde n_1,m_1\tilde m_1}(\lambda)$ and $A^{(3)}_{n_1\tilde n_1,m_1\tilde m_1}(\lambda)$, we need to examine $\partial_\lambda f_1$ and $\partial_\lambda\tilde f_1$, which can be generally decomposed as
\begin{eqnarray}
\partial_\lambda f_1(\lambda)&=&u(\lambda)f_1(\lambda)+u'(\lambda)f_1^\dag(\lambda)+v(\lambda)\tilde f_1(\lambda)+v'(\lambda)\tilde f_1^\dag(\lambda)+\sum_{\alpha}\bigr[a(\lambda)d_\alpha(\lambda)+a'(\lambda)d_{\alpha}^\dag(\lambda)\bigr]\nonumber\\
&&+\sum_{\alpha}\bigr[b(\lambda)\tilde d_\alpha(\lambda)+b'(\lambda)\tilde d_{\alpha}^\dag(\lambda)\bigr]\\
\partial_\lambda\tilde f_1(\lambda)&=&\tilde u(\lambda)\tilde f_1(\lambda)+\tilde u'(\lambda)\tilde f_1^\dag(\lambda)+\tilde v(\lambda)f_1(\lambda)+\tilde v'(\lambda)f_1^\dag(\lambda)+\sum_{\alpha}\bigr[\tilde a(\lambda)\tilde d_\alpha(\lambda)+\tilde a'(\lambda)\tilde d_{\alpha}^\dag(\lambda)\bigr]\nonumber\\
&&+\sum_{\alpha}\bigr[\tilde b(\lambda)d_\alpha(\lambda)+\tilde b'(\lambda)d_{\alpha}^\dag(\lambda)\bigr].
\end{eqnarray}
Fermion modes composed of other Majorana zero states are neglected in the above expansion since they are far away from $f_1$ and $\tilde f_1$. Note that the terms corresponding to bulk quasi-particle operators $d_\alpha$ and $\tilde d_\alpha$ cannot contribute to the Berry's connection. Therefore for simplicity we also neglect them in the further discussion.
From the time-reversal transformation of the two complex fermion operators ${\cal T}^{-1}f_1{\cal T}=\tilde f_1, {\cal T}^{-1}\tilde f_1{\cal T}=-f_1$, we have the following restrictions in the coefficients
\begin{eqnarray}
u(\lambda)=u^*(\lambda), \ u'(\lambda)=[u'(\lambda)]^*,\ v(\lambda)=-\tilde v^*(\lambda),\ v'(\lambda)=-[v'(\lambda)]^*.
\end{eqnarray}
The fermion parity conservation for each sector requires that $v(\lambda)=v'(\lambda)=0$. Therefore we have
\begin{equation}
\begin{split}
\partial_\lambda f_1(\lambda)&=u(\lambda)f_1(\lambda)+u'(\lambda)f_1^\dag(\lambda),\nonumber\\
\partial_\lambda\tilde f_1(\lambda)&=u^*(\lambda)\tilde f_1(\lambda)+u'^{*}(\lambda)\tilde f_1^\dag(\lambda).
\end{split}
\end{equation}
Substituting the above formulas into the second and third lines in Eqs.~\eqref{eqn:Berry0} we get then
\begin{equation}
\begin{split}
A^{(2)}_{n_1\tilde n_1,m_1\tilde m_1}(\lambda)&=i\hbar(1-m_1)u(\lambda)\delta_{n_1m_1}\delta_{\tilde n_1\tilde m_1},\\
A^{(3)}_{n_1\tilde n_1,m_1\tilde m_1}(\lambda)&=i\hbar(1-\tilde m_1)u^*(\lambda)\delta_{n_1m_1}\delta_{\tilde n_1\tilde m_1}.
\end{split}
\end{equation}
Furthermore, it can be shown that
\begin{eqnarray}
u(\lambda)\propto\langle\psi(\lambda)|f_1^\dag(\partial_\lambda f_1)|\psi(\lambda)\rangle=0,
\end{eqnarray}
where we have used the results that $\gamma_j^2(\lambda)=1$ and the overlapping between $\gamma_1$ and $\gamma_2$ is negligible. With these results in mind we have $A^{(2,3)}_{n_1\tilde n_1,m_1\tilde m_1}(\lambda)=0$, and conclude that
\begin{eqnarray}
A_{n_1\tilde n_1,m_1\tilde m_1}(\lambda)=i\hbar\langle\psi(\lambda)|\partial_\lambda |\psi(\lambda)\rangle\delta_{n_1m_1}\delta_{\tilde n_1\tilde m_1}\delta_{n_2m_2}\delta_{\tilde n_2\tilde m_2}...\delta_{n_Mm_M}\delta_{\tilde n_M\tilde m_M},
\end{eqnarray}
which is independent of $n_j$ and $\tilde n_j$. Since the Berry's connection is diagonal and identical for all qubit states, the Berry phase effect does not bring any nontrivial contribution to the braiding process. This result is similar as the situation in the chiral topological superconductors.

\subsection*{B.2. Ground state variation}

The nontrivial effect for the exchange of two pairs of Majorana zero modes can be resulted from the fact that each ground state itself $|n_1...n_M\rangle|\tilde n_1...\tilde n_M\rangle$ can change after the braiding process. Actually, the final state is generically related to the initial one via (after braiding $\gamma_{1},\tilde\gamma_1$ and $\gamma_2,\tilde\gamma_2$)
\begin{eqnarray}\label{eqn:ground2}
|n_1...n_M\rangle|\tilde n_1...\tilde n_M\rangle_{\rm final}=e^{i\theta(n_1,\tilde n_1)}|n_1n_2...n_M\rangle|\tilde n_1\tilde n_2...\tilde n_M\rangle_{\rm initial}.
\end{eqnarray}
We can always choose $\theta(1,1)=0$. Since the off-diagonal Berry's connection is zero, the above results show that braiding Majorana zero modes $\gamma_{1},\tilde\gamma_1$ and $\gamma_2,\tilde\gamma_2$ is equivalent to two independent processes of exchanging $\gamma_{1}$ and $\gamma_2$, and $\tilde\gamma_1$ and $\tilde\gamma_2$, respectively. For the complex fermion operators, we have $f_1(\lambda_{\rm final})=e^{i\theta(1,0)}f_1(\lambda_{\rm initial})$ and $\tilde f_1(\lambda_{\rm final})=e^{i\theta(0,1)}\tilde f_1(\lambda_{\rm initial})$. From the time-reversal symmetry we have that $\theta(1,0)=-\theta(0,1)$. This leads to
\begin{eqnarray}\label{eqn:ground3}
|1n_2...n_M\rangle|\tilde1\tilde n_2...\tilde n_M\rangle_{\rm final}&=&|1n_2...n_M\rangle|1\tilde n_2...\tilde n_M\rangle_{\rm initial},\\
|1n_2...n_M\rangle|\tilde0\tilde n_2...\tilde n_M\rangle_{\rm final}&=&e^{i\theta_0}|1n_2...n_M\rangle|0\tilde n_2...\tilde n_M\rangle_{\rm initial},\\
|0n_2...n_M\rangle|\tilde1\tilde n_2...\tilde n_M\rangle_{\rm final}&=&e^{-i\theta_0}|0n_2...n_M\rangle|1\tilde n_2...\tilde n_M\rangle_{\rm initial},\\
|0n_2...n_M\rangle|\tilde0\tilde n_2...\tilde n_M\rangle_{\rm final}&=&|0n_2...n_M\rangle|0\tilde n_2...\tilde n_M\rangle_{\rm initial},
\end{eqnarray}
where $\theta_0\equiv\theta(1,0)=-\theta(0,1)$. In the next subsection we shall prove that $\theta_0=\pi/2$. Then the braiding matrix is given by $U_{12}=e^{\frac{\pi}{4}\gamma_1\gamma_2}e^{\frac{\pi}{4}\tilde\gamma_1\tilde\gamma_2}$, which explicitly respects the time-reversal symmetry.

\subsection{Braiding phases}

Since the exchange dynamics is proved to be equivalent to two independent processes of braiding the Majorana zero modes $\gamma_{1},\tilde\gamma_1$ and $\gamma_2,\tilde\gamma_2$, respectively, we can construct a toy model of the DIII class Majorana quantum wire to study the braiding phases. The simplest case is a two-copy of 1D Kitaev model \cite{KitaevApp} with spin-$1/2$ fermions respecting the time-reversal symmetry. A T-junction is needed and can be formed by a vertical wire (along $y$ axis) which has $N$ sites, and a horizontal wire (along $x$ direction) which has $2N+1$ sites. The $N$th site of the vertical wire connects to the $N+1$th site of the vertical wire. Moreover, we consider that in the topological region of the T-junction, the hopping term equals to the pairing term $t=\Delta$. We model the Hamiltonian that
\begin{eqnarray}
H&=&-\mu\sum_{\sigma;y=1}^{N}c_{y,\sigma}^\dag c_{y,\sigma}+|t|\sum_{x=1}^{2N+1}\bigr[(e^{\phi}c_{x,\uparrow}^\dag+e^{-\phi}c_{x,\uparrow})(e^{\phi}c_{x+1,\uparrow}^\dag-e^{-\phi}c_{x+1,\uparrow})\nonumber\\
&&+(e^{-\phi}c_{x,\downarrow}^\dag+e^{\phi}c_{x,\downarrow})(e^{-\phi}c_{x+1,\downarrow}^\dag-e^{\phi}c_{x+1,\downarrow})\bigr],
\end{eqnarray}
where $\phi$ is the hopping and pairing phase for the vertical wire and $\mu<0$ is the chemical potential for the horizontal wire. The chemical potential for the horizontal wire, the hopping and pairing in the vertical wire are set to be zero. In this configuration the vertical wire of the T-junction is initially in the trivial phase and the horizontal wire in the topological phase. Note the braiding dynamics should be independent of $\phi$, we shall consider $\phi=0$ for simplicity, and then
\begin{eqnarray}\label{eqn:toy1}
H=-\mu\sum_{\sigma;y=1}^{N}c_{y,\sigma}^\dag c_{y,\sigma}+t\sum_{x=1}^{2N}\bigr[(c_{x+1,\uparrow}^\dag+c_{x+1,\uparrow})(c_{x,\uparrow}^\dag-c_{x,\uparrow})+
(c_{x+1,\downarrow}^\dag+c_{x+1,\downarrow})(c_{x,\downarrow}^\dag-c_{x,\downarrow})\bigr].
\end{eqnarray}

With the above model, we have four decoupled Majorana zero bound states, $\gamma_{1,\uparrow,\downarrow}$ and $\gamma_{2N+1,\uparrow,\downarrow}$, localized at two end sites. The initial four degenerate ground states are given by
\begin{eqnarray}
|1\rangle|\tilde1\rangle_{\rm initial}&=&\frac{1}{2^{2N}}\prod_{\sigma=\uparrow,\downarrow}\biggr[1+\sum_{p=0}^{N}\sum_{i_1<...<i_{2p}}^{2N+1}c_{i_{2p,\sigma}}^\dag...c_{i_{1},\sigma}^\dag \biggr]|\rm vac\rangle,\nonumber\\
|1\rangle|\tilde0\rangle_{\rm initial}&=&\frac{1}{2^{2N}}\biggr[1+\sum_{p=0}^{N}\sum_{i_1<...<i_{2p}}^{2N+1}c_{i_{2p},\uparrow}^\dag...c_{i_{1},\uparrow}^\dag\biggr] \biggr[\sum_{p=0}^{N}\sum_{i_1<...<i_{2p+1}}^{2N+1}c_{i_{2p+1},\downarrow}^\dag...c_{i_{1},\downarrow}^\dag\biggr]|\rm vac\rangle,\nonumber\\
|0\rangle|\tilde1\rangle_{\rm initial}&=&\frac{1}{2^{2N}}\biggr[\sum_{p=0}^{N}\sum_{i_1<...<i_{2p+1}}^{2N+1}c_{i_{2p+1},\uparrow}^\dag...c_{i_{1},\uparrow}^\dag\biggr] \biggr[1+\sum_{p=0}^{N}\sum_{i_1<...<i_{2p}}^{2N+1}c_{i_{2p},\downarrow}^\dag...c_{i_{1},\downarrow}^\dag\biggr]|\rm vac\rangle,\nonumber\\
|0\rangle|\tilde0\rangle_{\rm initial}&=&\frac{1}{2^{2N}}\prod_{\sigma=\uparrow,\downarrow}\sum_{p=0}^{N}\sum_{i_1<...<i_{2p+1},\sigma}^{2N+1}c_{i_{2p+1}}^\dag...c_{i_{1},\sigma}^\dag|\rm vac\rangle.\nonumber
\end{eqnarray}
It is straightforward to verify that for all above states the average number of electrons is $\bar N=2N+1$.

The Majorana end modes can be transported by tuning adiabatically the parameters $\mu$ and $t$ ($=\Delta$) in the T-junction \cite{AliceaApp}. After the braiding the Majorana modes $\gamma_{1,\uparrow,\downarrow}$ and $\gamma_{2N+1,\uparrow,\downarrow}$ exchange their positions, which can be pictorially described by reversing the pairing direction in the Hamiltonian~\eqref{eqn:toy1}. This implies that after braiding the new Hamiltonian is obtained by taking $\Delta\rightarrow-\Delta$, since the pairing to the left and right directions in the Hamiltonian explicitly has opposite sign. Therefore we get the new Hamiltonian (note now $t=-\Delta$)
\begin{eqnarray}\label{eqn:}
H'=-\mu\sum_{\sigma;y=1}^{N}c_{y,\sigma}^\dag c_{y,\sigma}-t\sum_{x=1}^{2N}\bigr[(c_{x,\uparrow}^\dag-c_{x,\uparrow})(c_{x+1,\uparrow}^\dag+c_{x+1,\uparrow})+
(c_{x,\downarrow}^\dag-c_{x,\downarrow})(c_{x+1,\downarrow}^\dag+c_{x+1,\downarrow})\bigr].
\end{eqnarray}
The corresponding new ground states then read
\begin{eqnarray}\label{eqn:}
|1\rangle|\tilde1\rangle_m&=&\frac{1}{2^{2N}}\prod_{\sigma=\uparrow,\downarrow}\biggr[1+\sum_{p=0}^{N}\sum_{i_1<...<i_{2p}}^{2N+1}(-1)^pc_{i_{2p,\sigma}}^\dag...c_{i_{1},\sigma}^\dag \biggr]|\rm vac\rangle,\nonumber\\
|1\rangle|\tilde0\rangle_m&=&\frac{1}{2^{2N}}\biggr[1+\sum_{p=0}^{N}\sum_{i_1<...<i_{2p}}^{2N+1}(-1)^pc_{i_{2p},\uparrow}^\dag...c_{i_{1},\uparrow}^\dag\biggr] \biggr[\sum_{p=0}^{N}\sum_{i_1<...<i_{2p+1}}^{2N+1}(-1)^pc_{i_{2p+1},\downarrow}^\dag...c_{i_{1},\downarrow}^\dag\biggr]|\rm vac\rangle,\nonumber\\
|0\rangle|\tilde1\rangle_m&=&\frac{1}{2^{2N}}\biggr[\sum_{p=0}^{N}\sum_{i_1<...<i_{2p+1}}^{2N+1}(-1)^pc_{i_{2p+1},\uparrow}^\dag...c_{i_{1},\uparrow}^\dag\biggr] \biggr[1+\sum_{p=0}^{N}\sum_{i_1<...<i_{2p}}^{2N+1}(-1)^pc_{i_{2p},\downarrow}^\dag...c_{i_{1},\downarrow}^\dag\biggr]|\rm vac\rangle,\nonumber\\
|0\rangle|\tilde0\rangle_m&=&\frac{1}{2^{2N}}\prod_{\sigma=\uparrow,\downarrow}\sum_{p=0}^{N}\sum_{i_1<...<i_{2p+1},\sigma}^{2N+1}(-1)^pc_{i_{2p+1}}^\dag...c_{i_{1},\sigma}^\dag|\rm vac\rangle.\nonumber
\end{eqnarray}
To finish the braiding, we need to adiabatically transform $H'$ back to the initial form $H$. This can be performed by considering the following Hamiltonian
\begin{eqnarray}\label{eqn:}
H_\lambda&=&-\mu\sum_{\sigma;y=1}^{N}c_{y,\sigma}^\dag c_{y,\sigma}-t\sum_{x=1}^{2N}\biggr[(e^{-i\lambda\pi/2}c_{x,\uparrow}^\dag-e^{i\lambda\pi/2}c_{x,\uparrow}) (e^{-i\lambda\pi/2}c_{x+1,\uparrow}^\dag+e^{i\lambda\pi/2}c_{x+1,\uparrow})+\nonumber\\
&&+(e^{i\lambda\pi/2}c_{x,\downarrow}^\dag-e^{-i\lambda\pi/2}c_{x,\downarrow})(e^{i\lambda\pi/2}c_{x+1,\downarrow}^\dag+e^{-i\lambda\pi/2}c_{x+1,\downarrow})\biggr],
\end{eqnarray}
where $\lambda$ is an adiabatic parameter. The adiabatic ground states are given by
\begin{eqnarray}\label{eqn:}
|1\rangle|\tilde1\rangle_{\lambda}&=&\frac{1}{2^{2N}}\prod_{\sigma=\uparrow,\downarrow}\biggr[1+\sum_{p=0}^{N} \sum_{i_1<...<i_{2p}}^{2N+1}(-1)^pe^{-i\sigma_z\lambda\pi p}c_{i_{2p,\sigma}}^\dag...c_{i_{1},\sigma}^\dag \biggr]|\rm vac\rangle,\nonumber\\
|1\rangle|\tilde0\rangle_{\lambda}&=&\frac{1}{2^{2N}}\biggr[1+\sum_{p=0}^{N}\sum_{i_1<...<i_{2p}}^{2N+1}(-1)^pe^{-i\lambda\pi p}c_{i_{2p},\uparrow}^\dag...c_{i_{1},\uparrow}^\dag\biggr] \biggr[\sum_{p=0}^{N}\sum_{i_1<...<i_{2p+1}}^{2N+1}(-1)^pe^{i\lambda\pi(p+1/2)}c_{i_{2p+1},\downarrow}^\dag...c_{i_{1},\downarrow}^\dag\biggr]|\rm vac\rangle,\nonumber\\
|0\rangle|\tilde1\rangle_{\lambda}&=&\frac{1}{2^{2N}}\biggr[\sum_{p=0}^{N}\sum_{i_1<...<i_{2p+1}}^{2N+1}(-1)^pe^{-i\lambda\pi(p+1/2)}c_{i_{2p+1},\uparrow}^\dag...c_{i_{1},\uparrow}^\dag\biggr] \biggr[1+\sum_{p=0}^{N}\sum_{i_1<...<i_{2p}}^{2N+1}(-1)^pe^{i\lambda\pi p}c_{i_{2p},\downarrow}^\dag...c_{i_{1},\downarrow}^\dag\biggr]|\rm vac\rangle,\nonumber\\
|0\rangle|\tilde0\rangle_{\lambda}&=&\frac{1}{2^{2N}}\prod_{\sigma=\uparrow,\downarrow}\sum_{p=0}^{N}\sum_{i_1<...<i_{2p+1},\sigma}^{2N+1}(-1)^pe^{-i\sigma_z\lambda\pi(p+1/2)} c_{i_{2p+1}}^\dag...c_{i_{1},\sigma}^\dag|\rm vac\rangle,\nonumber
\end{eqnarray}

From the former subsection we know already that the Berry's phase $\theta_b$ is the same for all states. With the above states we can check directly that $\theta_b=0$ for all states. The vanishing Berry's phase is because the time-reversal partners in each ground state contributes oppositely to the Berry's phase. This is reasonable, since a nonzero diagonal Berry's phase actually breaks time-reversal symmetry. For $\lambda=0$ we have $H(\lambda)=H'$ and for $\lambda=1$ the Hamiltonian transforms back to the initial one $H(\lambda=1)=H$. Therefore at $\lambda=1$ we obtain the final ground states by
\begin{eqnarray}
|1\rangle|\tilde1\rangle_{\rm final}&=&|1\rangle|1\rangle_{\rm initial},\label{eqn:phase1}\\
|1\rangle|\tilde0\rangle_{\rm final}&=&i|1\rangle|\tilde0\rangle_{\rm initial},\\
|0\rangle|\tilde1\rangle_{\rm final}&=&-i|0\rangle|\tilde1\rangle_{\rm initial},\\
|0\rangle|\tilde0\rangle_{\rm final}&=&|0\rangle|\tilde0\rangle_{\rm initial}.\label{eqn:phase2}
\end{eqnarray}
We therefore complete the proof that the braiding phase $\theta_0\equiv\theta(1,0)=-\theta(0,1)=\pi/2$.

\subsection{Braiding matrix and applications}

According to the results in Eqs.~\eqref{eqn:phase1} to~\eqref{eqn:phase2}, the braiding matrix for exchanging $\gamma_j,\tilde\gamma_{j}$ and $\gamma_{j+1},\tilde\gamma_{j+1}$ can be constructed by $U_{j,j+1}(T,\tilde T)=e^{\frac{\pi}{4}\gamma_j\gamma_{j+1}}e^{\frac{\pi}{4}\tilde\gamma_j\tilde\gamma_{j+1}}$, which explicitly respects the time-reversal symmetry. To visualize the non-Abelian statistics, we consider now two DIII Majorana chains, with the two pairs of Majorana zero modes $\gamma_1,\tilde\gamma_1$ and $\gamma_2,\tilde\gamma_2$ localized in the first chain, and the other two pairs $\gamma_3,\tilde\gamma_3$ and $\gamma_4,\tilde\gamma_4$ in the second chain. The four complex fermion modes are defined by
\begin{eqnarray}\label{eqn:complex3}
f_1=\frac{1}{2}(\gamma_1+i\gamma_2), f_2=\frac{1}{2}(\gamma_3+i\gamma4), \tilde f_1=\frac{1}{2}(\tilde\gamma_1-i\tilde\gamma_2), \tilde f_2=\frac{1}{2}(\tilde\gamma_3-i\tilde\gamma4).
\end{eqnarray}
They satisfy the relation ${\cal T}^{-1}f_{1,2}{\cal T}=\tilde f_{1,2}$ and ${\cal T}^{-1}\tilde f_{1,2}{\cal T}=-f_{1,2}$.
In terms of the complex fermion modes, the transformation matrix for braiding $\gamma_2,\tilde\gamma_2$ and $\gamma_3,\tilde\gamma_3$ takes the form
\begin{eqnarray}\label{eqn:braid1}
U_{23}(T,\tilde T)=\frac{1}{2}(1+if_2^\dag f_1^\dag-if_2^\dag f_1+i f_2f_1^\dag-i f_2f_1)(1-i\tilde f_2^\dag\tilde f_1^\dag+i\tilde f_2^\dag\tilde f_1-i\tilde f_2\tilde f_1^\dag+i\tilde f_2\tilde f_1).
\end{eqnarray}

The Hilbert space of the four complex fermions is spanned by sixteen topological qubit states $|n_1n_2\rangle|\tilde n_1\tilde n_2\rangle=(f_1^\dag)^{n_1}(f_2^\dag)^{n_2}(\tilde f_1^\dag)^{\tilde n_1}(\tilde f_2^\dag)^{\tilde n_2}|00\rangle|\tilde0\tilde0\rangle$, where $n_j,\tilde n_j=0,1$. The bases can be explicitly written down in the form $(|00\rangle,f_1^\dag|00\rangle,f_2^\dag|00\rangle,f_1^\dag f_2^\dag|00\rangle|)\otimes(|\tilde0\tilde0\rangle,\tilde f_1^\dag|\tilde0\tilde0\rangle,\tilde f_2^\dag|\tilde0\tilde0\rangle,\tilde f_1^\dag \tilde f_2^\dag|\tilde0\tilde0\rangle|)$. With this basis we have further
\begin{eqnarray}\label{eqn:braid2}
U_{23}(T,\tilde T)=\frac{1}{2}{\left[
\begin{matrix}
1 & 0 & 0 & -i\\
0 & 1 & -i & 0\\
0 & -i & 1 & 0\\
-i & 0 & 0 & 1\\\end{matrix} \right]}\otimes{\left[
\begin{matrix}
1 & 0 & 0 & i\\
0 & 1 & i & 0\\
0 & i & 1 & 0\\
i & 0 & 0 & 1\\\end{matrix} \right]}.
\end{eqnarray}
Using the above braiding matrix and for an arbitrary initial state we can obtain the final state straightforwardly. If the initial state is $|00\rangle|\tilde0\tilde0\rangle$, for instance, we get
\begin{eqnarray}\label{eqn:braid3}
U_{23}(T,\tilde T)|00\rangle|\tilde0\tilde0\rangle&=&\frac{1}{2}\bigr(|00\rangle|\tilde0\tilde0\rangle+|11\rangle|\tilde1\tilde1\rangle +i|00\rangle|\tilde1\tilde1\rangle-i|11\rangle|\tilde0\tilde0\rangle\bigr)\nonumber\\
&=&\frac{1}{2}\bigr(|0\tilde0\rangle_L|0\tilde0\rangle_R+|1\tilde1\rangle_L|1\tilde1\rangle_R +i|1\tilde0\rangle_L|1\tilde0\rangle_R-i|0\tilde1\rangle_L|0\tilde1\rangle_R\bigr),
\end{eqnarray}
where the indices $L$ and $R$ represent the left and right Majorana chains, respectively.
It is interesting that the above state is generically a four-particle entangled state, which carries rich quantum information depending on different measurement strategies (see the discussion on measurement in the next section).
First, for each Majorana wire if we do not distinguish the states of the same total parity (e.g. $|1\tilde0\rangle_L$ and $|0\tilde1\rangle_L$) in the measurement, in the right hand side of the above state the former two terms are equivalent, and can be denoted as $|L_{\rm even}\rangle|R_{\rm even}\rangle$, which implies that both Majorana wires are in the even parity state. Similarly, the later two terms are also equivalent, and can be denoted by $|L_{\rm odd}\rangle|R_{\rm odd}\rangle$, implying that both Majorana wires are in the odd parity state. With these notations one can reduce the original state into an effective two-qubit entangled one
\begin{eqnarray}\label{eqn:braid5}
U_{23}(T,\tilde T)|00\rangle|\tilde0\tilde0\rangle=\frac{1}{\sqrt{2}}\bigr(|L_{\rm even}\rangle|R_{\rm even}\rangle+|L_{\rm odd}\rangle|R_{\rm odd}\rangle\bigr),
\end{eqnarray}
On the other hand, to identify the state~\eqref{eqn:braid1} as a four-particle entangled one, we should be able to measure the fermion parity for each sector of the time-reversal partners in a single wire. A novel scheme for the measurement will be proposed and studied in the next section. Similarly, with three Majorana wires of DIII class, we can generate a six-qubit code entanglement through two braiding processes. This shows the natural advantage in generating multi-particle entangled state using DIII class topological superconductors, which can be very useful in the quantum information processing. For example, a five-qubit code entanglement is the minimum requirement to realize an error correcting code \cite{ECC1,ECC2}.

Furthermore, a full braiding, i.e. braiding twice $\gamma_2,\tilde\gamma_2$ and $\gamma_3,\tilde\gamma_3$ yields the final state by
\begin{eqnarray}\label{eqn:braid6}
U_{23}^2(T,\tilde T)|00\rangle|\tilde0\tilde0\rangle=|1\tilde1\rangle|1\tilde1\rangle,
\end{eqnarray}
which distinguishes from the initial state in that each copy of the $p$-wave superconductor changes fermion parity. In contrast, a full braiding of two pairs of Majorana fermions in a chiral topological superconductor transforms the state back to the initial one, which is therefore always trivial. From these discussions we find that in the braiding the Majorana modes $\gamma_j$ do not feel their time-reversal partners $\tilde \gamma_j$, which is an essential difference from the situation in exchanging two pairs of Majorana fermions in a chiral superconductor, and makes the braiding operator in a TRI topological superconductor nontrivial.


\noindent

\end{document}